\documentclass[11pt,letterpaper]{article}
\usepackage{graphicx}
\usepackage[margin=1in]{geometry}
\usepackage{amsmath}
\usepackage{makecell}
\usepackage{hyperref}
\usepackage[font=small,labelfont=bf]{caption}
\usepackage{subcaption}

\title{PSPACE-Completeness of Reversible Deterministic Systems}
\author{
  Erik D. Demaine%
    \thanks{MIT Computer Science and Artificial Intelligence Laboratory, 32 Vassar Street, Cambridge, MA 02139, USA, \protect\url{{edemaine,dylanhen}@mit.edu}}
\and
  Robert A. Hearn\thanks{\protect\url{bob@hearn.to}}
\and
  Dylan Hendrickson\footnotemark[1]
\and
  Jayson Lynch\thanks{University of Waterloo Cheriton School of Computer Science, Waterloo, ON, Canada, \protect\url{jayson.lynch@uwaterloo.ca}}
}
\date{}

\newtheorem{theorem}{Theorem}

\graphicspath{{./figures/}}

\begin{document}
\maketitle

\begin{abstract}
  We prove PSPACE-completeness of several reversible, fully deterministic
  systems.
  At the core, we develop a framework for such proofs
  (building on a result of Tsukiji and Hagiwara and
  a framework for motion planning through gadgets),
  showing that any system that can implement three basic gadgets
  is PSPACE-complete.
  We then apply this framework to four different systems,
  showing its versatility.
  First, we prove that Deterministic Constraint Logic is PSPACE-complete,
  fixing an error in a previous argument from 2008.
  Second, we give a new PSPACE-hardness proof for the reversible `billiard ball'
  model of Fredkin and Toffoli from 40 years ago, newly establishing hardness
  when only two balls move at once.
  Third, we prove PSPACE-completeness of zero-player motion planning with
  any reversible deterministic interacting $k$-tunnel gadget and a
  `rotate clockwise' gadget (a zero-player analog of branching hallways).
  Fourth, we give simpler proofs that zero-player motion planning is
  PSPACE-complete with just a single gadget, the 3-spinner.
  These results should in turn make it even easier to prove PSPACE-hardness
  of other reversible deterministic systems.
\end{abstract}

\section{Introduction}

Reversible deterministic systems arise in various situations, some of the most
important of which come from physics because fundamental existing physical theories are reversible and deterministic\footnote{The time evolution of the wave-function in the Standard Model is deterministic even if the observation of macroscopic phenomena is probabilistic.}.  In particular, due to the
thermodynamics of information, reversible computation can potentially use significantly less energy than irreversible computation because Landauer's Principle requires physical systems expend $k_B T \ln 2$ energy per bit of information lost.%
\footnote{Here $k_B \approx 1.4 \cdot 10{-23}$ is the Boltzmann constant
  and $T$ is the temperature in kelvins.  At room temperature, this comes to
  about $2.8 \cdot 10^{-21}$ joules per bit.  Current chips are rapidly
  approaching this limit; see \cite{frank2020fundamental,Energy_ITCS2016}.}
Thus understanding how reversible systems can solve computationally difficult
problems may help in designing general-purpose reversible computing hardware.

More precisely, a system is \emph{deterministic} if its configuration at each
time in the future is entirely determined by its current configuration.
A system is \emph{reversible} if, in addition, its configuration at each time
in the past is entirely determined by its current configuration.
The systems we consider all satisfy, or nearly satisfy, the stronger property
of \emph{time-reversal symmetry}: evolution forward in time and backward in time
obey the same rules, so by looking at a sequence of configurations it is not
possible to determine whether time is moving forwards or backwards. To reverse
time, we simply need to reverse the direction of motion of each moving part in each of the systems we consider. In one system, we use a slightly more general symmetry by replacing each `rotate
clockwise' gadget with a `rotate counterclockwise' gadget, and vice-versa.
A physicist might call this parity--time (PT) symmetry; see, e.g.,
\cite{PT-symmetry}.

In this paper (Section~\ref{sec:framework}), we develop a framework for
proving PSPACE-completeness of reversible deterministic systems.
Our framework extracts and simplifies a framework implicit in the work of
Tsukiji and Hagiwara \cite{tsukiji2011recognizing}, who proved PSPACE-hardness
for Langton's reversible `ant' model of artificial life in two geometries,
the square and hexagonal grids.
Their hardness reductions construct five core gadgets in each grid,
and show that these gadgets suffice for PSPACE-hardness by a reduction from
satisfiability in Quantified Boolean Formulas (QBF).
Our framework decreases the number of required gadgets to just three,
showing that some of the previous gadgets are unnecessary (essentially,
redundant) and others can be simplified.
The framework also guarantees that the gadgets are connected together without
crossings, making it well suited to reducing to planar systems
(which all of our applications are).

We then apply our framework to analyzing the complexity of
four reversible deterministic systems:
\begin{enumerate}
\item
  We prove in Section~\ref{sec:dcl} that Deterministic Constraint Logic
  is PSPACE-complete.  While this result was already claimed 14 years ago
  \cite{CL_Complexity2008,hearn2009games}, we describe in 
  Section~\ref{sec:dcl error} an error in the previous reduction.
  The new framework enables a correct proof of the same result.
\item
  We develop in Section~\ref{sec:billiard} a new PSPACE-hardness proof for
  the `billiard ball' reversible model of computation, introduced and analyzed
  by Fredkin and Toffoli in 1982 \cite{fredkin1982conservative}.
  In this model, unit-radius 2D balls move without friction and collide
  elastically with pinned or movable objects, according to classical physics.
  Unlike the previous proof, our PSPACE-hardness result works
  even in the case when only two balls ever move at once
  (and the rest are stationary), which results in a substantially simpler proof
  (no longer needing complex timing arguments to guarantee simultaneity).
\item
  We prove in Section~\ref{sec:locking 2-toggles} that zero-player motion
  planning through gadgets is PSPACE-complete when the gadgets include
  \emph{any} reversible deterministic interacting $k$-tunnel gadget
  and a `rotate clockwise' gadget (a 1-state 3-location gadget where
  an entering signal simply exits along the clockwise-next location).
  This result can be thought of as extending Table~1 in the
  motion-planning-through-gadgets framework \cite{demaine2020toward}
  to add a `zero-player' column in the unbounded row,
  analogous to zero-player Deterministic Constraint Logic \cite{hearn2009games}.
  Our proof indeed uses the same simulations as for motion planning
  with a positive number of players \cite{demaine2020toward} to reduce to
  one core case --- locking 2-toggles and rotate clockwise ---
  and then shows that that case is PSPACE-complete.
\item
  We prove in Section~\ref{app:3-spinners} that zero-player motion planning
  with one very simple gadget called a `3-spinner' is PSPACE-complete.
  Specifically, a \emph{3-spinner} has two states --- `clockwise' and
  `counterclockwise' --- and three locations at which the signal can enter;
  after entering, the gadget flips its state and the signal exits in the
  next port in the order given by the state.
  This result is weaker than Tsukiji and Hagiwara's PSPACE-hardness of `ant'
  on a hexagonal lattice \cite{tsukiji2011recognizing}, because the vertices
  in the lattice act exactly as 3-spinners.
  We effectively translate this result into the motion-planning-through-gadgets
  framework of Demaine et al.~\cite{demaine2020toward}, and simplify it
  significantly.
\end{enumerate}

All of the systems we consider can straightforwardly be simulated using polynomial space, so the decision problems are in PSPACE.

\section{The Framework}\label{sec:framework}

Our framework for proving PSPACE-hardness, which is a modest simplification of one due to Tsukiji and Hagiwara \cite{tsukiji2011recognizing}, can be understood in terms of the motion-planning gadgets framework of Demaine et al.~\cite{demaine2020toward}. In particular, it is closely related to, and can be described in terms of, the `input/output gadgets' of Ani et al.~\cite{ani2020trains}. We will describe it independently.

The framework may apply to any setting with a single \emph{signal} deterministically navigating a planar \emph{network} of \emph{gadgets} with the following properties. Each gadget has some designated \emph{ports}. When the signal enters the gadget at one of its ports, it then exits the same gadget at one of its port, which is determined by the entrance port and any previous traversals of that gadget. The network links gadgets by connecting the ports of the gadgets in disjoint pairs: when the signal exits at a port, it enters at the paired port.

To describe the ``behavior'' of a gadget, we define a \emph{traversal} to be of the form $a \to b$ for any two ports $a$ and $b$ of the gadget.
A gadget \emph{implements} a sequence $[a_1\to b_1,\dots,a_k\to b_k]$ of traversals if, when the sequence of the signal's entrance ports to the gadget is $[a_1, \dots, a_k]$, the sequence of exit ports from the gadget is $[b_1, \dots, b_k]$. Note that a gadget implements any prefix of a sequence it implements.

All of the gadgets we consider in this section are \emph{symmetric under time-reversal}, meaning if we perform a sequence of traversals followed by its time-reverse, the gadget is returned to its original state. Formally, if a gadget implements two sequences $X=[a_1\to b_1,\dots,a_k\to b_k]$ and $Y=[c_1\to d_1,\dots c_\ell\to d_\ell]$, then it also implements
$$XX^{-1}Y=[a_1\to b_1,\dots,a_k\to b_k,b_k\to a_k,\dots,b_1\to a_1,c_1\to d_1,\dots,c_\ell\to d_\ell].$$
In the language of Hendrickson \cite{hendrickson2021gadgets}, our gadgets can be modeled as `prefix-closed gizmos', and time-reversal symmetry means they satisfy the `implication property' $X,Y\implies XX^{-1}Y$.

If every gadget in a network is symmetric under time-reversal, then the entire network is as well: if we reverse the direction of the signal by returning it to the just-exited port instead of the port paired to just-exited port, it will retrace its steps in reverse, eventually returning to the initial configuration. This is a special case of a result applying to implication properties in general \cite{hendrickson2021gadgets}.

\subsection{Required Gadgets}

We are now ready to describe the gadgets which we will show suffice for PSPACE-hardness. 

We describe each gadget by specifying some sequences it implements. The gadgets then also implement all prefixes of implemented sequences, and all sequences required for time-reversal symmetry. We don't fully specify the behavior of the gadgets: they are allowed to do anything if the signal arrives in an unspecified sequence, and this does not affect our PSPACE-hardness result because it never happens in the networks created by the reduction. The required behavior of our gadgets is summarized in Table~\ref{fig:gadgets summary}.
In addition, for each gadget $G$ described below, we also allow our network to include the gadget \emph{$G$ after $[\alpha_1\to\beta_1, \dots, \alpha_i\to\beta_i]$}, which behaves like $G$ would after having performed the traversals $\alpha_1\to\beta_1, \dots, \alpha_i\to\beta_i$ in that order.
That is, if $G$ implements $[\alpha_1 \to \beta_1, \dots, \alpha_i \to \beta_i, a_1 \to b_1, \dots, a_k \to b_k]$, then $G$ after$[\alpha_1\to\beta_1, \dots, \alpha_i\to\beta_i]$ implements $[a_1\to b_1,\dots,a_k\to b_k]$.

Our first, and most complicated gadget, is the \emph{Switch}. This corresponds to three of Tsukiji and Hagiwara's gadgets, the `Switch \& Pass,' `Switch \& Turn,' and `Pseudo-Crossing,' which are all equivalent except for the cyclic order of ports in the planar embedding, and that Switch \& Turn merges the ports we call Set and Out. The Switch has 5 ports, called `Set,' `Out,' `Test,' `T-Out,' and `F-Out.' It implements $[\text{Set}\to\text{Out}, \text{Test}\to\text{T-Out}]$ and $[\text{Test}\to\text{F-Out}]$. Intuitively, it has an internal state which is initially False, and is set to True by the traversal $\text{Set}\to\text{Out}$. Entering Test reveals the current state. Time-reversal symmetry implies that the Switch is reusable: for instance, it must also implement
$$[\text{Set}\to\text{Out}, \text{Test}\to\text{T-Out}, \text{T-Out}\to\text{Test}, \text{Out}\to\text{Set}, \text{Test}\to\text{F-Out}].$$

There are really 12 different Switch gadgets (up to rotation and reflection), based on the cyclic order of the ports. We allow any cyclic order of the ports; our PSPACE-hardness applies to any individual order.

Our next gadget is the \emph{Reversible Fan-in}. Tsukiji and Hagiwara call this gadget `CONJ.' It has three ports $a$, $b$, and $c$, and implements $[a\to c]$ and $[b\to c]$. Intuitively, it is a fan-in that sends both $a$ and $b$ to $c$, but---as required by time-reversal symmetry---remembers which entrance was taken so that when the signal returns to $c$, it exits the port it originally entered.

Our final gadget is the \emph{A/BA Crossover}. The A/BA Crossover has four ports $A$, $B$, $a$, and $b$ in cyclic order, and implements $[A\to a]$ and $[B\to b,A\to a]$. Tsukiji and Hagiwara build a slightly more powerful crossover they call `CROSS,' which also implements $[A\to a,B\to b]$. However, this is not necessary for PSPACE-hardness, and the A/BA Crossover can easily be constructed using Tsukiji and Hagiwara's Pseudo-Crossing (which is a particular planar embedding of a Switch) and CONJ.

\begin{table}
  \centering
  \begin{tabular}{|c|c|c|c|}\hline
    Gadget & Ports & Cyclic Order & Implements \\\hline\hline
    Switch & \makecell[l]{Set \\ Out \\ Test \\ T-Out \\ F-Out} & Any order & \makecell[l]{$[\text{Set}\to\text{Out}, \text{Test}\to\text{T-Out}]$\\ $[\text{Test}\to\text{F-Out}]$}\\\hline
    Reversible Fan-in & \makecell[l]{$a$\\$b$\\$c$} & (Only one possible) & \makecell[l]{$[a\to c]$\\$[b\to c]$}\\\hline
    A/BA Crossover & \makecell[l]{$A$\\$B$\\$a$\\$b$} & $A$, $B$, $a$, $b$ & \makecell[l]{$[A\to a]$\\$[B\to b,A\to a]$}\\\hline
  \end{tabular}
  \caption{Summary of time-reversal-symmetric gadgets required for PSPACE-hardness. Each gadget implements all sequences generated from those under Implements by prefixes and time-reversal symmetry ($X,Y\implies XX^{-1}Y$).}
  \label{fig:gadgets summary}
\end{table}

\subsection{PSPACE-Hardness}

We now prove PSPACE-hardness for the natural decision problem concerning these gadgets: given a planar network containing Switches, Reversible Fan-ins, and A/BA Crossovers (including these gadgets after some traversals), a starting port which the signal enters first, and a target port, does the signal ever reach the target port? We reduce from QBF, still following Tsukiji and Hagiwara \cite{tsukiji2011recognizing} with some simplification and slightly different abstractions.

We first ignore the requirement of planarity, showing PSPACE-hardness for general networks containing just Switches and Reversible Fan-ins. Then we argue that A/BA Crossovers suffice for all required crossings in a planar embedding of the networks we construct.

\begin{figure}
  \centering
  \includegraphics[scale=.7]{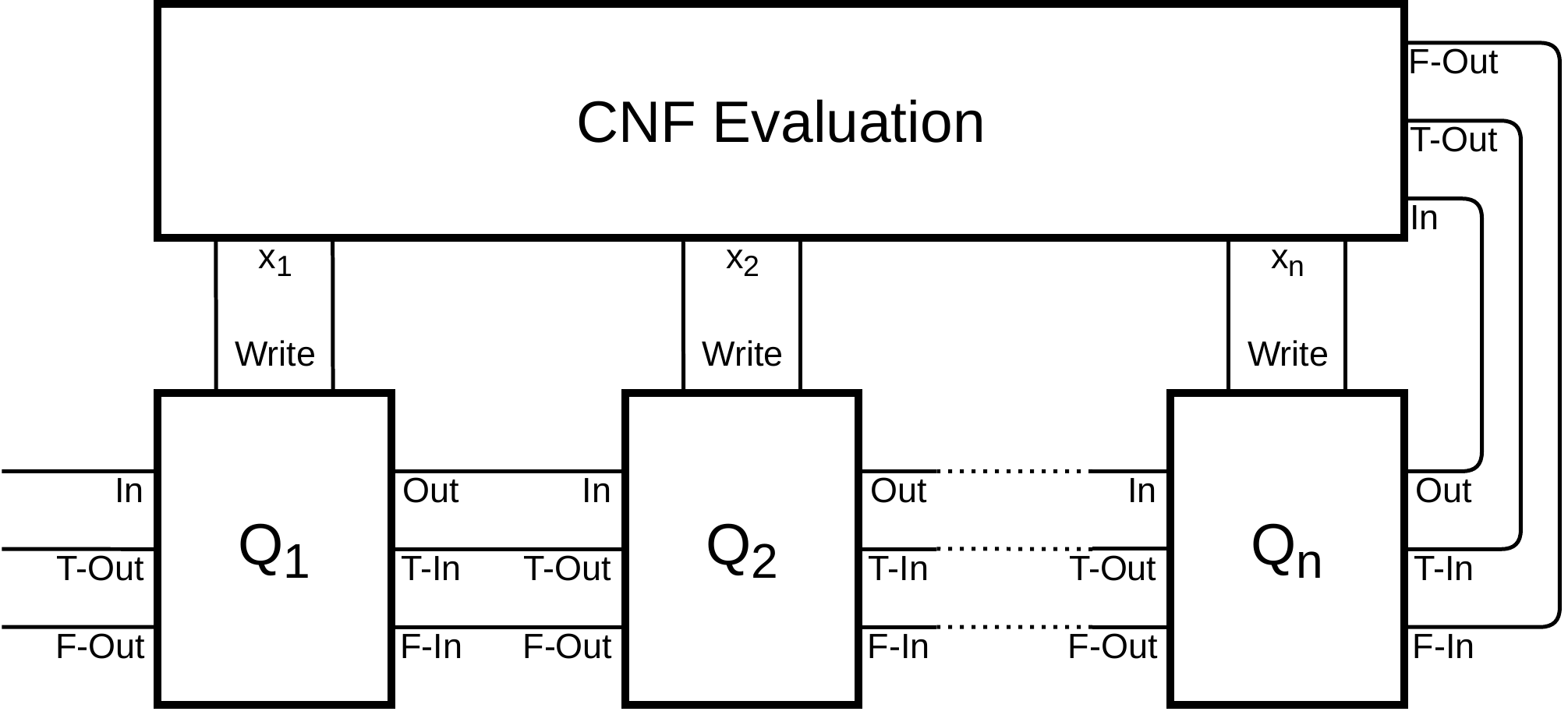}
  \caption{The high-level structure of the network produced by our reduction. The signal begins at In on $Q_1$, evaluates the formula, and eventually arrives at T-Out or F-Out on $Q_1$ depending on its truth value.}
  \label{fig:reduction overview}
\end{figure}

Given a quantified formula $Q_1x_1:\cdots Q_nx_n:\phi(x_1,\dots,x_n)$ where $\phi$ is a 3-CNF formula, we construct a network of Switches and Reversible Fan-ins. At a high level, the network consists of a series of `quantifier gadgets,' ending in `CNF evaluation.' When the signal arrives at a quantifier gadget, the quantifier gadget sets the variable it controls, and then queries the next quantifier. Depending on the response, it may perform a second query with the other setting of its variable, and then it sends a response to the previous quantifier. The final quantifier $Q_n$ instead queries the CNF evaluation, which computes the value of $\phi$ under the current variable assignment. The structure of the reduction is shown in \figurename~\ref{fig:reduction overview}.

Because we are working with gadgets which are symmetric under time-reversal, we need our quantifier gadgets have this symmetry as well. Quantifiers need to be used multiple times, so we will reset them in the way suggested by time-reversal symmetry: the signal needs to backtrack across its entire path through each quantifier gadget before returning to the previous quantifier. We will describe the desired behavior of quantifier gadgets which are symmetric under time-reversal, and later show how to build them using Switches and Reversible Fan-ins.

We specifically discuss universal quantifiers; existential quantifiers require only a minor modification. A universal quantifier gadget $Q_i$ has eight locations, named in cyclic order `F-Out', `T-Out', `In', `Write-Out', `Write-In', `Out', `T-In', and `F-In.'%
\footnote{Tsukiji and Hagiwara call these `OUT$_{i,\text{FALSE}}$,' `OUT$_{i,\text{TRUE}}$,' `IN$_i$,' `I$_{x_i}$,' `O$_{x_i}$,' `IN$_{i+1}$,' `OUT$_{i+1,\text{TRUE}}$,' and `OUT$_{i+1,\text{FALSE}}$,' respectively.}
The gadget is activated when the signal arrives at In, and the signal proceeds to Out to query the next quantifier; the variable $x_i$ is currently set to False.

Eventually, the signal returns at either T-In or F-In, indicating the truth value of the remainder of the formula with the current variable assignment up to $x_i$. If it enters at F-In, the universally quantified formula is false, so it passes this along to $Q_{i-1}$ by exiting at F-Out. If it enters at T-In, we need to try the other assignment, which means we need to reset the quantifiers after $Q_i$ by backtracking through them. So the quantifier gadget `remembers' that it received one True signal, and sends the signal back out T-In. Due to reversibility, the signal eventually returns to Out, at which point it is sent to Write-Out to set $x_i$ to True. The signal goes through a series of Switches in the CNF evaluation, and then returns at Write-In. Now $Q_i$ sends the signal to Out, this time with the other setting of $x_i$. Eventually the signal returns again at either T-In or F-In, and it is sent straight to T-Out or F-Out to answer the query from $Q_{i-1}$.

Once $Q_{i-1}$ has dealt with the response, the signal returns to $Q_i$ at the same one of T-Out or F-Out it exited, at which point everything is reversed, ending with the signal exiting at In with $x_i$ set to False, and $Q_i$ and all later quantifiers in their initial configuration.

Formally, we need a universal quantifier to implement these sequences (and those implied by time-reversal symmetry), corresponding to the first query to $Q_{i+1}$ returning False, the first query returning True but the second returning False, and both queries returning True, respectively:
\begin{itemize}
  \item $[\text{In}\to\text{Out}, \text{F-In}\to\text{F-Out}]$
  \item $[\text{In}\to\text{Out}, \text{T-In}\to\text{T-In}, \text{Out}\to\text{Write-Out},\text{Write-In}\to\text{Out}, \text{F-In}\to\text{F-Out}]$
  \item $[\text{In}\to\text{Out}, \text{T-In}\to\text{T-In}, \text{Out}\to\text{Write-Out},\text{Write-In}\to\text{Out}, \text{T-In}\to\text{T-Out}]$
\end{itemize}
An existential quantifier gadget is constructed by swapping T-In with F-In and T-Out with F-Out on a universal quantifier gadget.

The signal starts at In on $Q_1$, which queries the truth value of the whole formula. It eventually arrives at either T-Out or F-Out depending on the answer; we make T-Out on $Q_1$ the target port. If we connect In, T-Out, and F-Out to themselves, then after evaluating the formula the signal will backtrack all the way to the beginning, and repeat this cycle.

The final quantifier $Q_k$ interfaces directly with the CNF evaluation instead of another quantifier. The CNF evaluation maintains the current variable assignment, initially with all variables False. It has a path for each variable $x_i$ which is connected to Write-Out and Write-In on $Q_i$; traversing this path forwards sets $x_i$ True, and then traversing it backwards returns $x_i$ to False. The CNF evaluation has three additional ports In, T-Out, and F-Out, analogous to those on a quantifier gadget. When the signal arrives at In, it exits at either T-Out or F-Out depending on the truth value of the formula under the current variable assignment. These ports are connected to Out, T-In, and T-Out on $Q_k$ in the same way as other quantifiers.

By the designed behavior of quantifier gadgets and CNF evaluation, the signal arrives at T-Out on $Q_1$ if and only if the quantified formula is true. We still need to fill in the details: how do we build quantifier gadgets and CNF evaluation and of Switches and Reversible Fan-ins, and how do we handle crossings?

\subsubsection{CNF evaluation}

\begin{figure}
  \centering
  \includegraphics[scale=.68]{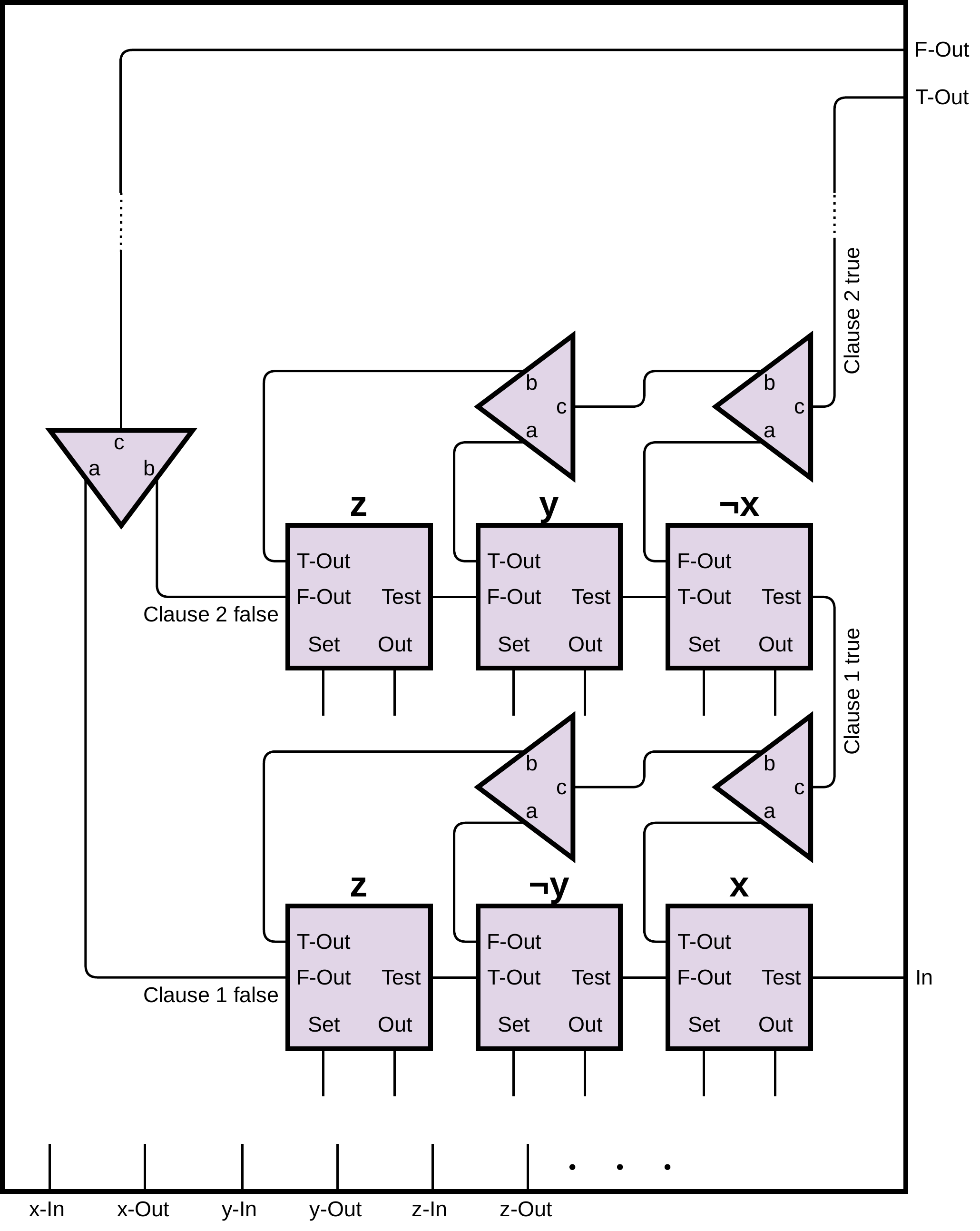}
  \caption{Our CNF evaluation. Each clause consists of three Switches corresponding to the literals in the clause, with Reversible Fan-ins to merge paths. A variable and its negation differ in the positions of T-Out and F-Out on the corresponding Switch. When the signal enters In, if any literal in the first clause is true it will take the edge labeled ``Clause 1 true'' and otherwise will take the edge labeled ``Clause 1 false.'' All the exits for false clauses merge and lead to F-Out. If all clauses are true, the signal will traverse them in series and then exit T-Out. For each variable $x_i$, there is also a path from $x_i$-In to $x_i$-Out which goes through $\text{Set}\to\text{Out}$ on the switch corresponding to each instance of $x_i$ or $\neg x_i$.}
  \label{fig:cnf evaluation}
\end{figure}

Our CNF evaluation is the same as Tsukiji and Hagiwara's, and is shown in \figurename~\ref{fig:cnf evaluation}. There is a switch for each literal in $\phi$. For each variable $x_i$, there is a path that goes through all switches corresponding to instances of $x_i$ (or $\neg x_i$) in $\phi$, and traversing this path sets $x_i$ to True. When the signal enters In, it checks each clause in series. For each clause, it goes through the switches corresponding to literals in the clause, and emerges in one of two locations depending on whether the clause is satisfied. If it is not satisfied, the signal exits at F-Out, and otherwise it proceeds to the next clause, exiting at T-Out once it has passed every clause. Later, it will return to either T-Out or F-Out and reverse its path back to In; the Reversible Fan-ins remember the path taken and necessarily send it back along the same path.

\subsubsection{Quantifier gadgets}

\begin{figure}
	\begin{minipage}[b]{0.58\textwidth}
  \centering
  \includegraphics[scale=.7]{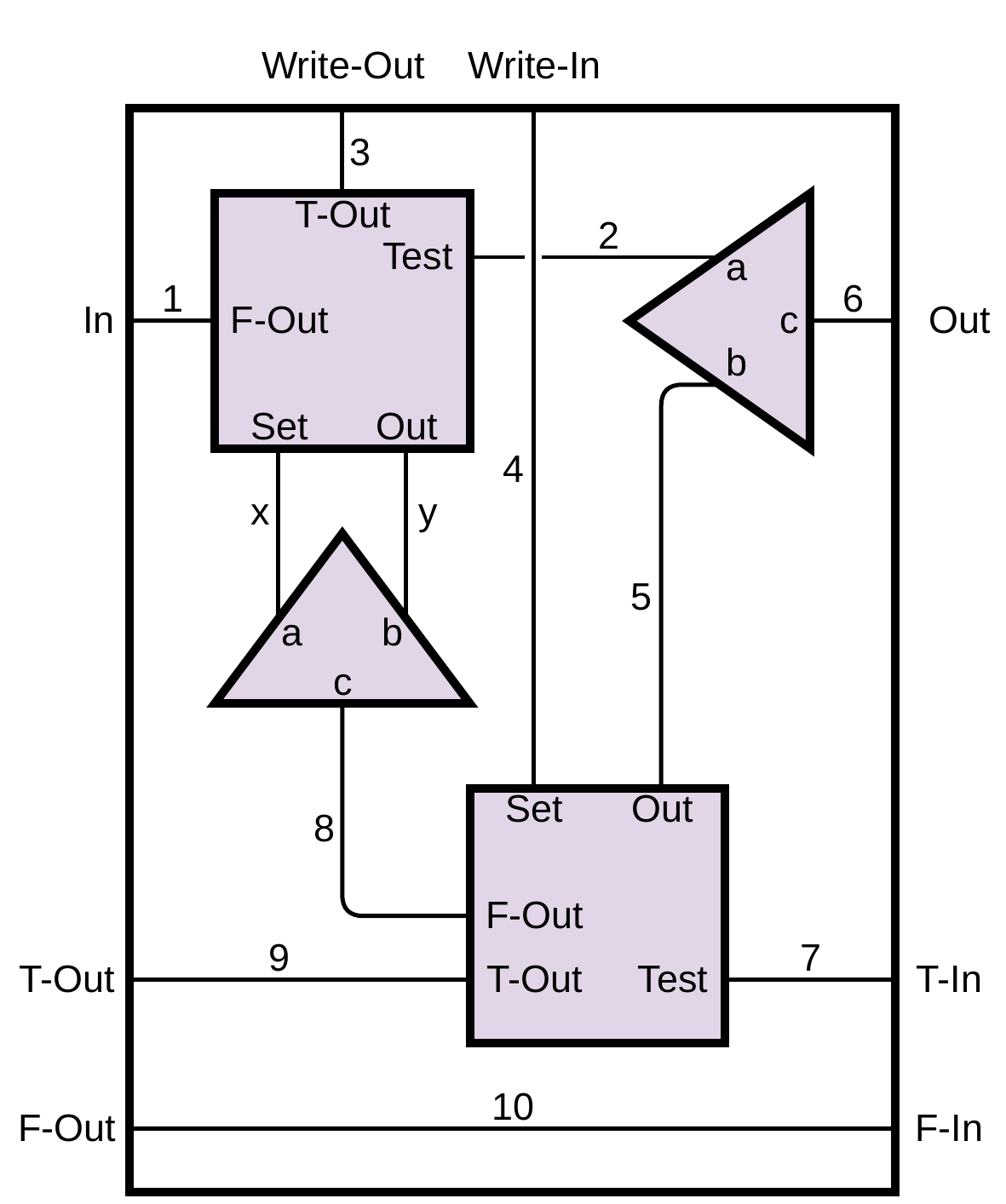}
  \caption{The universal quantifier gadget built from two Switches (squares) and two Reversible Fan-ins (triangles). The top Switch begins is after $[\text{Test}\to\text{F-Out}]$, the bottom left Reversible Fan-in is after $[a\to c]$, and the other two gadgets are their default versions. Edges between gadgets are labeled for later use.}
  \label{fig:universal quantifier}
  \end{minipage}
\hfill
	\begin{minipage}[b]{0.35\textwidth}
	\centering
	\includegraphics{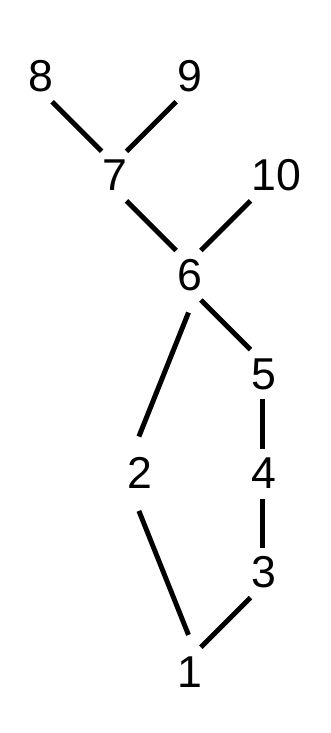}
	\caption{A Hasse diagram of the order relation on used edges in our quantifier gadgets (\figurename~\ref{fig:universal quantifier}). That $a$ is above $b$ indicates that whenever both edges are used, $a$ was used more recently and will be unused sooner.}
	\label{fig:hasse}
	\end{minipage}
\end{figure}

Our quantifier gadgets are essentially the same as Tsukiji and Hagiwara's, the only differences are due to planar arrangement and that we must build their Switch \& Turn gadget out of a Switch and a Reversible Fan-in. The universal quantifier gadget is shown in \figurename~\ref{fig:universal quantifier}. The existential quantifier gadget is constructed by exchanging the roles of T-In with F-In and T-Out with F-Out, so there is a direct path from T-In to T-Out which crosses some edges linking F-In and F-Out to the other ports. This similarity is sensible: for existential quantifiers if the formula is false we need to try again with the other value, but for universal quantifiers if the formula is true we are allowed to attempt the other required value for the variable.

We must check that the universal quantifier gadget correctly implements the behavior described above. Recall that the signal will first arrive at In. It proceeds to the upper left switch, taking $[\text{F-Out}\to\text{Test}]$ and leaving the Switch in its default state. Then signal takes $[a\to c]$ in the upper right Fan-in and leaves at Out. If it now enters F-In, it goes directly to F-Out. If instead it enters T-In, it goes from Test to F-Out on the bottom switch, goes along edge $8$ to the Reversible Fan-In (which is after $[a\to c]$), and traverses $[c\to a]$. Then the signal traverses $\text{Set}\to\text{Out}$ on the top switch, and returns to the bottom switch via the Reversible Fan-in, leaving both the Switch and Reversible Fan-in in different states than before. The signal then backtracks from F-Out to Test on the bottom Switch, and exits T-In, where it just entered. Now if the signal enters Out, the Reversible Fan-in sends it back to Test on the top Switch along edge 2. But the top Switch has been activated, so the signal exits the Switch at T-Out and exits the quantifier at Write-Out. It next enters Write-In, at which point it traverses $\text{Set}\to\text{Out}$ on the bottom Switch, and exits Out. Finally, if the signal now enters F-In, it is still sent to F-Out, and if it enters T-In then it goes from Test to T-Out on the bottom switch (which has now been activated) and exits the quantifier at T-Out.

\subsubsection{Planarity}

Finally, we argue that we can use A/BA Crossovers to avoid crossings in the network produced by this reduction. 

Note that each edge in the network is directed, in the sense that the first traversal across the edge is in a predetermined direction which we call \emph{forwards}, and all future traversals alternate direction---we never traverse an edge twice consecutively in the same direction. At any time while running the system, we say an edge is \emph{used} if it has been traversed forwards more recently than backwards. Initially no edges are used, and they are used and unused throughout the process. For two edges $x$ and $y$ which cross, an A/BA Crossover suffices for their crossing provided that whenever both $x$ and $y$ are used, always the same edge---say $x$---was traversed forwards more recently, and also $x$ will be traversed backwards sooner in the future. In this case, we can set $x$ to be the $A\to a$ tunnel and $y$ to be the $B\to b$ tunnel of an A/BA Crossover. If $x$ and $y$ are never both used, either orientation of the A/BA Crossover will work.

So we just need to argue that there is a consistent order edges (other than the few we showed can avoid crossings) are used. There are no crossings outside the CNF evaluation and quantifier gadgets, so we need only check those gadgets. For the CNF evaluation, this is straightforward:
\begin{itemize}
  \item For $i<j$, the path to set $x_i$ is used before the path to set $x_j$.
  \item Within the path to set $x_i$, the edges are used in order.
  \item All paths for setting variables are used before edges involved in testing the current value.
  \item The edges involved in testing the current value are used in order. Specifically, there is a partial order on these edges based on when it is possible to traverse one and then another on the way from In to T-Out or F-Out. We arbitrarily extend this partial order to a total order, or equivalently, for two edges which can't both be used, we arbitrarily choose which is A and which is B in the A/BA Crossover.
\end{itemize}

For quantifier gadgets, the numbering listed in \figurename~\ref{fig:universal quantifier} works as an order for all edges other than $x$ and $y$. More generally, a Hasse diagram of the ``is sometimes used after'' partial order on these edges is shown in \figurename~\ref{fig:hasse}, and positioning A/BA Crossovers to respect this order suffices for all crossings between these edges. It is straightforward to verify this partial order by considering the behavior of our quantifier gadgets.

For crossings inside a quantifier gadget which involve edge $x$ or $y$, we need a different approach: for instance, if edge $2$ crosses $x$, then the signal will sometimes traverse $2$, then $x$, then $2$ backwards, which isn't supported by the default A/BA Crossover. When $x$ or $y$ is involved in crossing, we use an A/BA crossover as follows:
\begin{itemize}
  \item If $x$ crosses $y$, make $x$ the $B\to b$ tunnel since it is always used first.\footnote{Alternatively, avoid this crossing by adjusting the Reversible Fan-in connecting $x$ and $y$.}
  \item If $x$ or $y$ crosses $1$, make $1$ the $B\to b$ tunnel since it is always used first.
  \item If $x$ or $y$ crosses $3$, $4$, $5$, $9$, or $10$, make $x$ or $y$ the $B\to b$ tunnel since they are always used first.
  \item If $x$ or $y$ crosses $2$, $6$, $7$, or $8$, use an A/BA Crossover after $A\to a$, and make the numbered tunnel $a\to A$. By time-reversal symmetry, the A/BA Crossover after $A\to a$ implements $[a\to A,B\to b,A\to a]$, which corresponds for instance to traversing $2$ forwards, $x$ backwards, and then $2$ backwards, which is what is needed.
\end{itemize}

To carefully check that this arrangement of A/BA Crossovers works for the quantifier gadget, we can consider the possible sequences of edge traversals. Using $\cdot^{-1}$ for backwards traversals, these are (generated by time-reversal symmetry from)
\begin{itemize}
  \item $[1,2,6,10]$
  \item $[1,2,6,7,8,x,y,8^{-1},7^{-1},6^{-1},2^{-1},3,4,5,6,10]$
  \item $[1,2,6,7,8,x,y,8^{-1},7^{-1},6^{-1},2^{-1},3,4,5,6,7,9]$
\end{itemize}
which correspond to the sequences the quantifier gadget was built to implement. It is straightforward to verify, for each pair of edges, that an A/BA Crossover as described supports all of the ways that pair of tunnels is used. For instance, the possible sequences for just $2$ and $x$ are $[2]$ and $[2,x,2^{-1}]$, which are $[a\to A]$ and $[a\to A,B\to b,A\to a]$ on the A/BA Crossover involved, and both of these are implemented by an A/BA Crossover after $A\to a$. It suffices to check just the sequences listed, since taking the closure under time-reversal symmetry does not give rise to any new intermediate configurations.

Hence we have the main result of this section:

\begin{theorem}
  Given a planar network of Switches, Reversible Fan-ins, A/BA Crossovers, and these gadgets after some traversals, a starting location, and a target location, it is PSPACE-complete to determine whether the signal ever reaches the target location from the starting location. This result holds even when all Switches have any particular cyclic order of ports.
\end{theorem}

To apply this framework to a specific problem, we simply need to describe the signal and how it moves along wires, and then construct a Switch (with ports in any order), Reversible Fan-in, and A/BA Crossover.

\section{Deterministic Constraint Logic}\label{sec:dcl}

Constraint Logic is a problem about graph orientation reconfiguration introduced by Hearn and Demaine \cite{CL_Complexity2008,hearn2009games} as a tool for proving hardness results. A \emph{constraint graph} is a directed planar graph where each edge has \emph{weight} 1 or 2, which are colored red and blue, respectively.%
\footnote{In grayscale, blue edges are darker than red edges. Figures also draw blue edges thicker than red edges.}
Each vertex in a constraint graph is either an AND vertex, which has two red and one blue edge, or an OR vertex, which has three blue edges. Each vertex is required to have at least 2 total weight in edges pointing towards it. Edges change orientation, while maintaining this constraint. Hearn and Demaine show how to `tie up' loose edges, allowing the use of degree-2 vertices with any combination of colors, for which the required weight is only 1 (so a single red edge satisfies it).

In this paper, we are specifically interested in \emph{Deterministic Constraint Logic (DCL)}, in which edges flip according to the following deterministic rule. Each time step, an edge flips if it didn't flip in the previous time step and it can flip without violating the in-weight constraint of the vertex it is currently directed towards, or it did flip in the previous time step but no other edge pointing towards the vertex it is now directed towards can flip this time step.

Here are the basic behaviors that result from the deterministic rule:
\begin{itemize}
  \item Begin with a path of edges of any color, all pointing to the left. If the leftmost edge flips, all the edges in the path will flip, one in each time step.
  \item If a blue edge flips to point towards an OR vertex, in the next time step the blue edge which was already pointing towards the OR vertex will flip.
  \item If a blue edge flips to point towards an AND vertex, in the next time step both red edges pointing towards that vertex will flip.
  \item If both red edges flip to point towards an AND vertex in the same time step, in the next time step the blue edge will flip.
  \item If one red edge but not the other flips to point towards an AND vertex, in the next time step the same red edge will flip again.
\end{itemize}

The decision problem in Deterministic Constraint Logic is whether some specified edge will eventually flip, given a constraint graph and the set of edges that are considered to have flipped in time step 0.

\subsection{Issue with Existing Proof}\label{sec:dcl error}

Hearn and Demaine's proof of PSPACE-hardness for Deterministic Constraint Logic \cite{hearn2009games} has a subtle issue. When their universal quantifier receives a `satisfied in' signal, it records this fact, much like our universal quantifier gadget. When it receives a second `satisfied in' signal (assuming the signal did not enter `try out' in between), it erases the record of the first one; this is by design, to reset the gadget for the next variable assignment.

The existential quantifier tries assigning its variable False, then True, and then False again, and passes every `satisfied in' signal it gets to `satisfied out' to inform the previous quantifier.
If the existential quantifier is satisfied when its variable is False but not True, it sends two such signals instead of one. This is the problem: if the previous quantifier is universal, the second signal cancels the first one, and that quantifier behaves as though there was no signal. The simplest formula for which the reduction fails is $\forall x\exists y:\neg y$. Modifying the existential quantifier to test each assignment exactly once does not fix the problem, because then if the quantifier is satisfied by both values for its variable, it sends two signals to the previous quantifier. In particular, $\forall x\exists y:y\lor\neg y$ would fail.

The proof may be fixable by modifying the existential quantifier gadget to ensure it only ever sends one signal; it would likely be about as complicated as the universal quantifier. The approach our framework takes is different: it adds an additional query return line, so instead of just `satisfied in' we have both T-in and F-in, and quantifier gadgets are guaranteed to receive exactly one response for each query.

\subsection{PSPACE-Hardness}

Our PSPACE-hardness proof for Deterministic Constraint Logic uses many of the same elements as Hearn and Demaine's. The signal is a flipping edge, which propagates along paths in the direction opposite the orientation of the edges in the path. Like Hearn and Demaine, our gadgets will sometimes contain `bouncing' edges which flip in a periodic way, and we ensure the length of each path through a gadget is a multiple of this period---for us, the period is 2, though Hearn and Demaine used a period of 4. The ports of our gadgets are always blue edges, which are connected by joining them with a degree-2 vertex. The target edge is the edge corresponding to the target port, and it flips if and only if the signal reaches the target port.

While DCL itself is symmetric under time reversal, it is possible to build a DCL gadget which is not, by including periodically bouncing edges calibrated such that the signal enters out of phase with when it exits. Some of Hearn and Demaine's gadgets \cite{hearn2009games} behave this way. However, all of our gadgets will be symmetric under time reversal in all of their relevant behavior, as is required for the framework we are using.

\begin{figure}
  \centering
  \includegraphics[scale=.4]{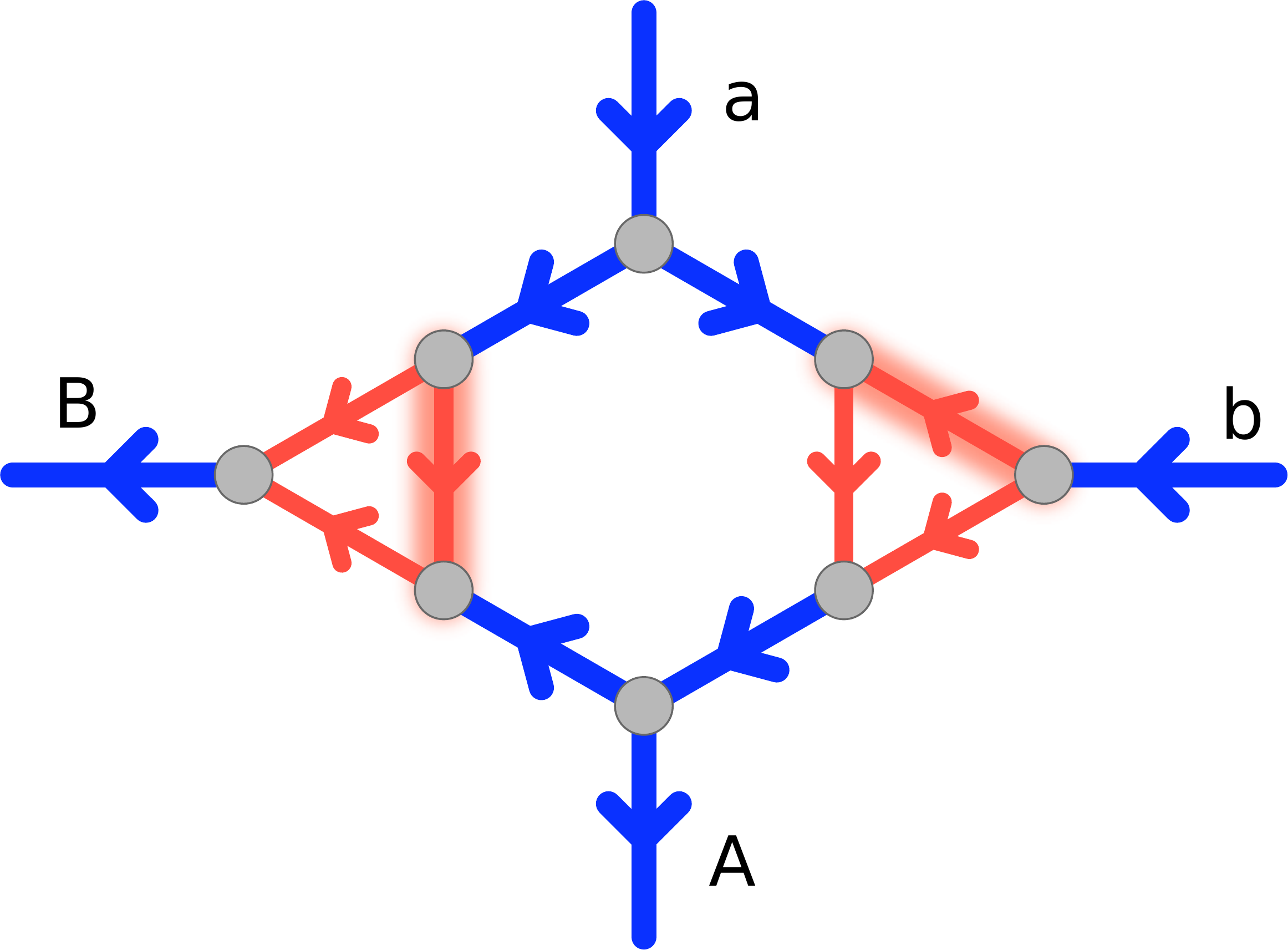}
  \caption{An A/BA Crossover for Deterministic Constraint Logic, from Hearn and Demaine \cite{hearn2009games}. Glowing auras indicate edges that flip every time step---the state shown is the state immediately before the signal enters the gadget, so that when the signal enters, the blue edge at the entered port and all glowing edges simultaneously flip from the shown configuration.}
  \label{fig:dcl crossover}
\end{figure}

We simply need to build valid Switch, Reversible Fan-in, and A/BA Crossover gadgets. A Reversible Fan-in is simply an OR vertex, which always takes 2 time steps to traverse. We use Hearn and Demaine's A/BA crossover, which we reproduce in \figurename~\ref{fig:dcl crossover}. This A/BA crossover always takes an even number of time steps to traverse, and contains bouncing edges with period 2.

\begin{figure}
  \centering
  \includegraphics[scale=.4]{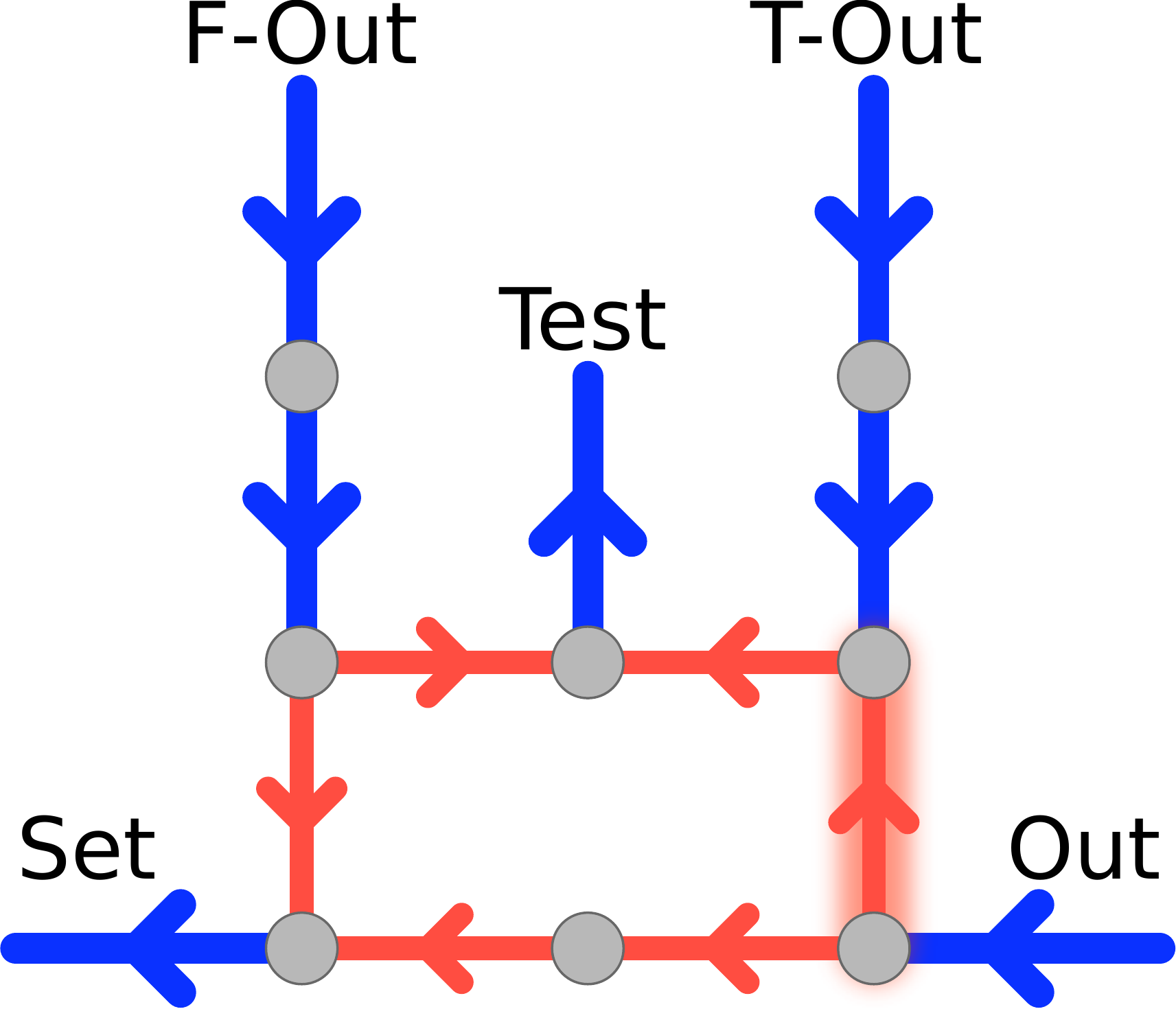}\hfil
  \includegraphics[scale=.4]{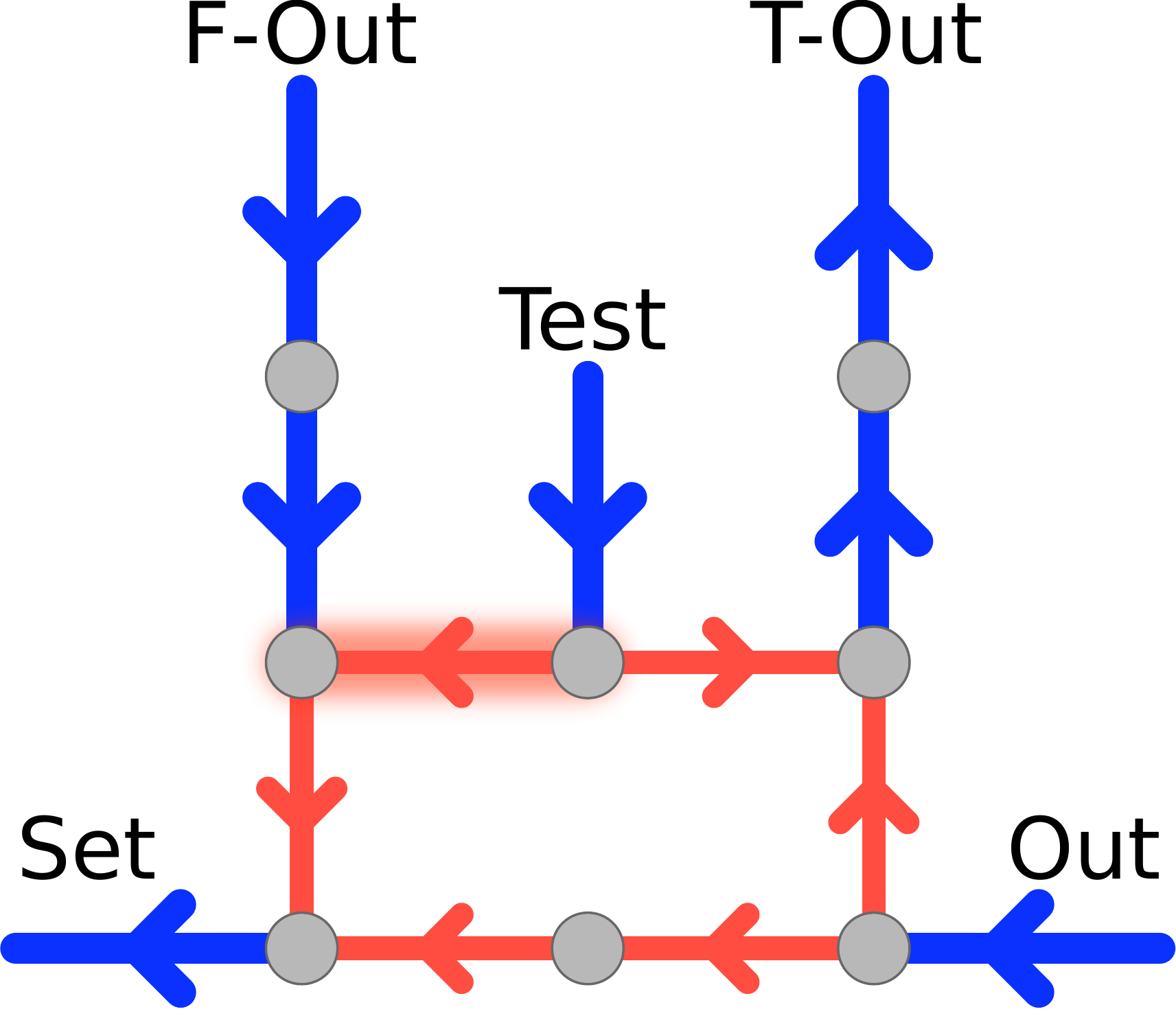}
  \caption{A Switch for Deterministic Constraint Logic. Left: the initial configuration. Right: the configuration after the traversal $\text{Test}\to\text{T-Out}$.}
  \label{fig:dcl switch}
\end{figure}

Our Switch gadget is a bit more complicated, and is shown in \figurename~\ref{fig:dcl switch}. If the signal arrives at Set, it exits at Out and reflects the configuration by flipping the bottom four edges and setting the left red edge bouncing instead of the right red edge. If the signal arrives at Test, it exits either F-Out or T-Out based on which red edge is currently bouncing, and sets one of the top red edges bouncing. Every traversal through this gadget takes four time steps.

\section{Locking 2-Toggles}\label{sec:locking 2-toggles}

\begin{figure}
  \centering
  \includegraphics{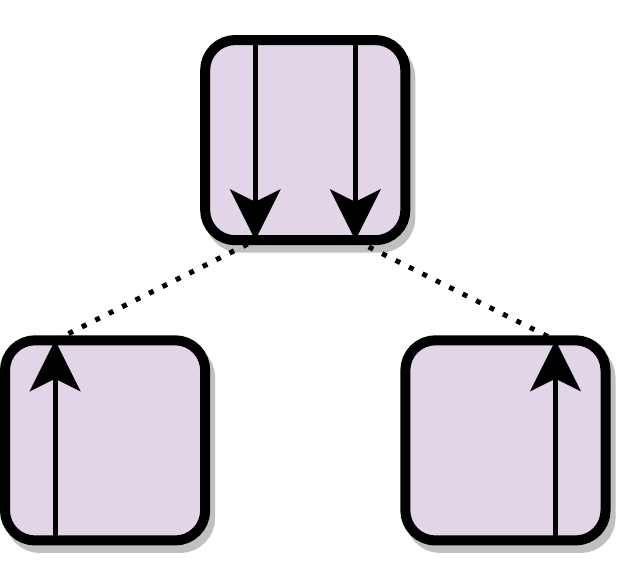}
  \caption{The locking 2-toggle. Dotted lines indicate state transitions upon the signal traversing a tunnel. All the locking 2-toggles we show have this layout, which Demaine et al.~\cite{demaine2020toward} call the `parallel' locking 2-toggle (and Demaine et al.'s results imply that any single planar embedding is sufficient).}
  \label{fig:l2t}
\end{figure}

Our next application is a zero-player version of decision problems considered by Demaine et al.~\cite{demaine2020toward}, and also fits in the framework from Section~\ref{sec:framework}. The \emph{locking 2-toggle} is a gadget with two directed tunnels, where traversing either one flips its direction and disables the other tunnel until the traversed tunnel is traversed again in the opposite direction. A diagram is shown in \figurename~\ref{fig:l2t}. To adapt the locking 2-toggle to the fully deterministic setting, we say that if the signal arrives at a port where it cannot currently cross the tunnel, it `bounces off,' exiting the same port. The locking 2-toggle is symmetric under time-reversal.

The locking 2-toggle alone is boring in this setting: because it has separate tunnels, the signal is restricted to a linear path. Since the locking 2-toggle obeys time-reversal symmetry, when the signal bounces it will backtrack everything it has done, and not produce interesting new behavior. In particular, the reachability question can be solved in logarithmic space.

To make a more interesting problem, we introduce a zero-player analog of the `branching hallways' used in one-player motion planning. This is a new gadget we call \emph{rotate clockwise}, and the other enantiomer \emph{rotate counterclockwise}. Rotate clockwise has three ports, and the signal always exits immediately clockwise of the port it entered. This is like a 3-spinner which doesn't change state. Rotate Clockwise and the notion of bouncing off of closed ports also appear in Asynchronous Ballistic Reversible Logic \cite{frank2017asynchronous}, which is similar to the model just defined, but with signals.

Rotate clockwise is not symmetric under time-reversal: the time-reversed rotate clockwise is exactly rotate counterclockwise. But we will build gadgets which are symmetric under time-reversal---at least, provided the sequence of input ports is one we need to account for---out of rotate clockwise and locking 2-toggles.

\begin{figure}
  \centering
  \includegraphics{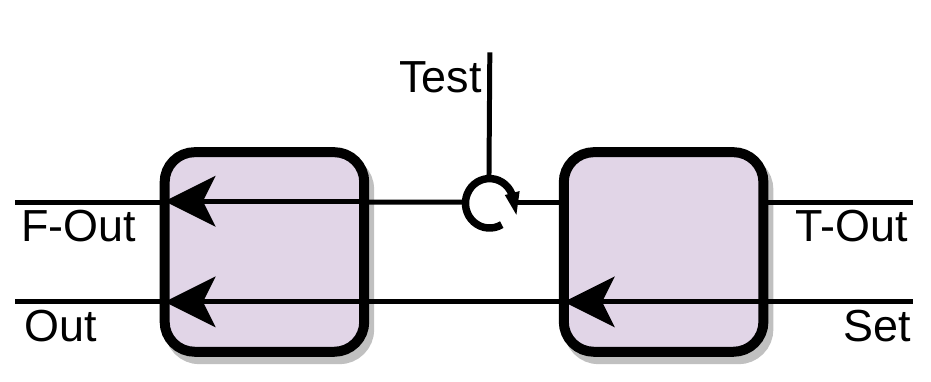}\hfil
  \includegraphics{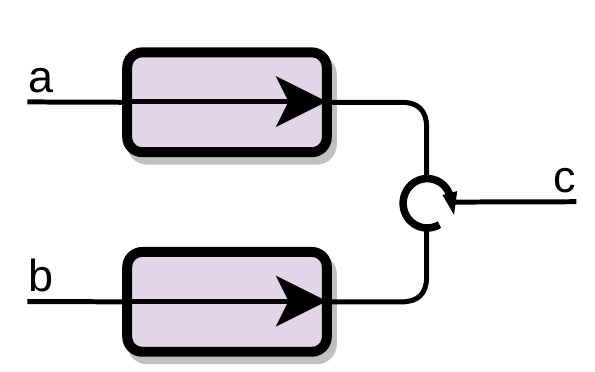}
  \caption{The Switch and Reversible Fan-in built out of locking 2-toggles and rotate clockwise. It is easy to verify correctness. Note that we cannot rely on time-reversal symmetry; we must check each desired sequence followed by its time-reverse, and this needs to return the gadget to its initial state. The A/BA Crossover from Demaine et al.~\cite{demaine2020toward} is complicated to build with parallel locking 2-toggles, so we omit it.}
  \label{fig:l2t gadgets}
\end{figure}

The decision problem is whether, in a given network of locking 2-toggles and rotate clockwise, the signal ever reaches some location. For PSPACE-hardness, we need to construct a Switch, a Reversible Fan-in, and an A/BA Crossover. These gadgets are shown in \figurename~\ref{fig:l2t gadgets}. The A/BA Crossover is the same as the one built by Demaine et al.~\cite{demaine2020toward} for the nondeterministic setting, with branching hallways replaced by rotate clockwise. We are able to construct all of these gadgets with just rotate clockwise, without using rotate counterclockwise.

Demaine et al.\ \cite{demaine2020toward} showed that any gadget from a large class---what they call interacting-tunnels reversible deterministic gadgets---can simulate the locking 2-toggle. These simulations do not use any branching hallways, so they work in our fully deterministic model as well. Hence zero-player motion planning with any interacting-tunnels reversible deterministic gadget and rotate clockwise is PSPACE-complete.

Our gadgets for locking 2-toggles, including those from Demaine et al. involved in building the A/BA Crossover, don't rely on the precise behavior of rotate clockwise: any instances of rotate clockwise could be replaced with rotate counterclockwise, and the reduction would still work. In fact---though this is more complicated to verify because the resulting gadgets are nondeterministic (in particular, the player can choose to turn around at any time, but this is the only resulting nondeterminism)---all instances of rotate clockwise can be replaced with branching hallways, yielding a reduction to one-player motion planning with locking 2-toggles. This is a new and arguably simpler proof of PSPACE-hardness than the original by Demaine et al.~\cite{demaine2020toward}.

\section{3-Spinners}\label{app:3-spinners}

One additional application of our framework is a strictly more general decision problem, and thus a strictly weaker hardness result, than the hexagonal-lattice version of Langton's ant which Tsukiji and Hagiwara prove PSPACE-complete \cite{tsukiji2011recognizing}. We include it because it resolves a question posed by Demaine et al.~\cite{demaine2018computational}, who were not aware of Tsukiji and Hagiwara's work: this result implies that 1-player motion planning with 3-spinners is PSPACE-complete, since the player would never have nontrivial decisions to make because 3-spinners obey time-reversal symmetry. We are able to simplify the gadgets involved a bit because we will not restrict to a hexagonal lattice like Tsukiji and Hagiwara do.

\begin{figure}
	\centering
	\includegraphics{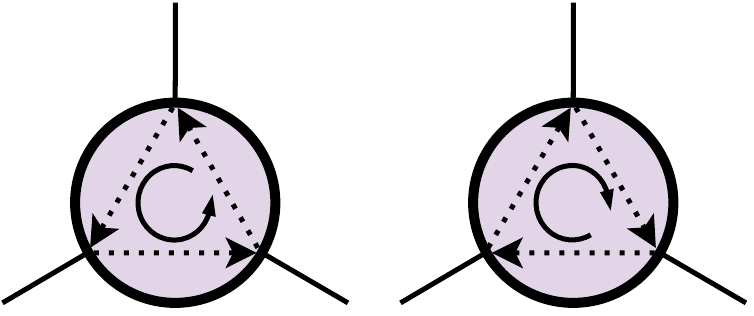}
	\caption{The 3-spinner. In the left state, the signal is sent one port counterclockwise, and in the right state it is sent one port clockwise, as indicating by the dotted arrows. Each traversal flips the state. The circular arrow indicates the current state.}
	\label{fig:3-spinner}
\end{figure}

The \emph{3-spinner} is a particular gadget which fits in the framework described in Section~\ref{sec:framework} and is symmetric under time-reversal. It has three locations $a_1$, $a_2$, and $a_3$, and implements all sequences which alternate between traversals of the form $a_i\to a_{i+1}$ and $a_i\to a_{i-1}$ with indices mod 3. That is, the first time the signal enters a 3-spinner, it exits one position `clockwise' of the entrance, the next time it exits one position `counterclockwise,' and this alternates. A diagram of the 3-spinner is shown in \figurename~\ref{fig:3-spinner}.

\begin{figure}
	\centering
	\includegraphics[scale=.7]{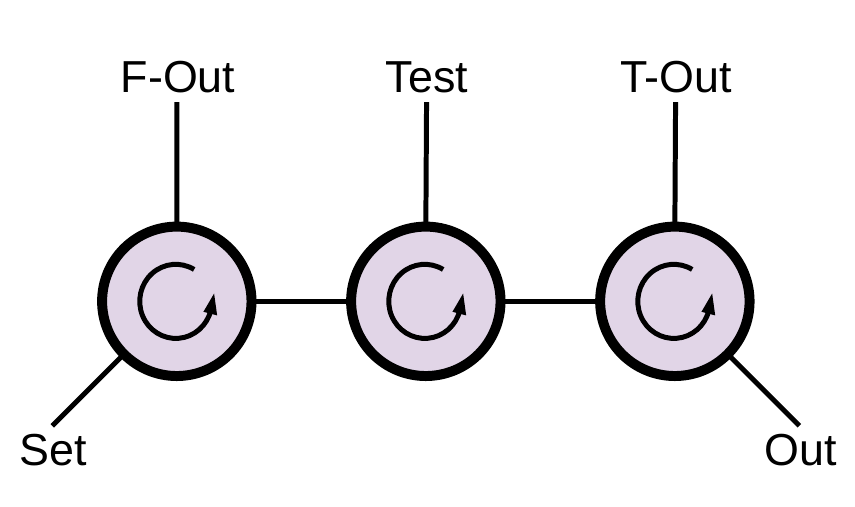}\hfill
	\includegraphics[scale=.7]{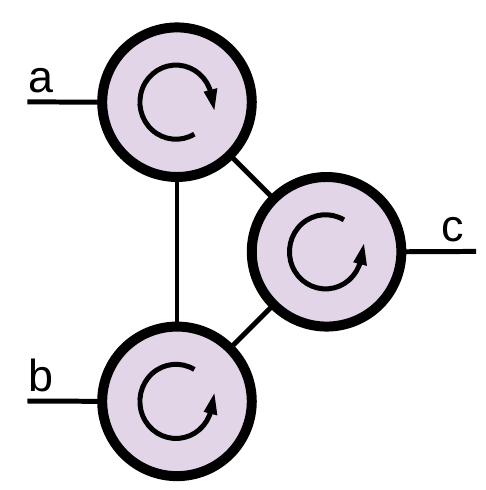}\hfill
	\includegraphics[scale=.7]{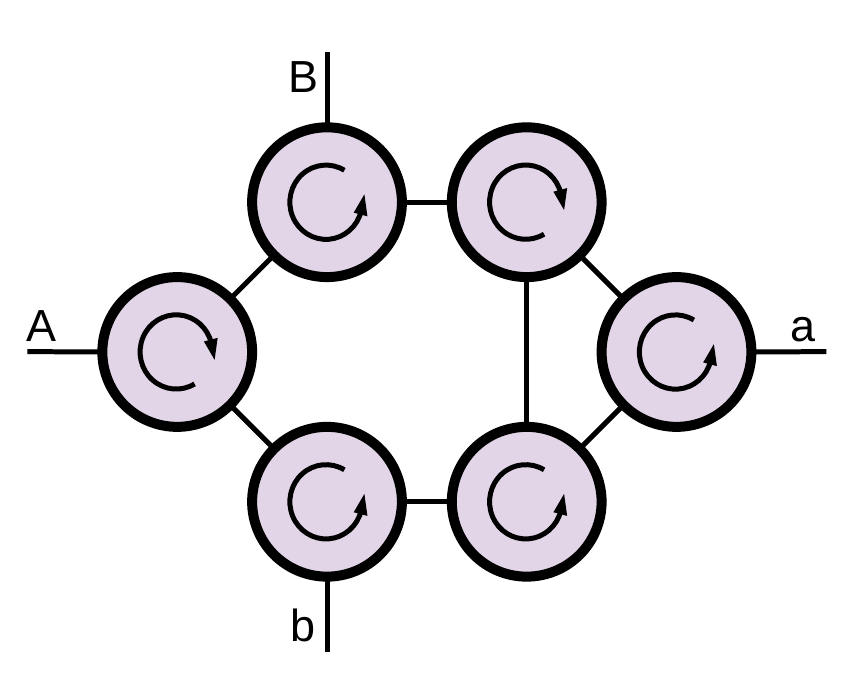}
	\caption{The Switch, Reversible Fan-in, and A/BA Crossover built out of 3-spinners, based on gadgets from Tsukiji and Hagiwara \cite{tsukiji2011recognizing}. Correctness is easily verified by testing each desired sequence of input ports. The A/BA Crossover is the combination of a (differently laid out) Switch and a Reversible Fan-in.}
	\label{fig:3-spinner gadgets}
\end{figure}

Zero-player motion planning with 3-spinners asks whether the signal ever reaches some location in a network of 3-spinners. To prove PSPACE-hardness even in planar networks, we just need to show how to build a Switch, a Reversible Fan-in, and an A/BA Crossover out of 3-spinners. These constructions are simplified versions of gadgets by Tsukiji and Hagiwara. Our gadgets are shown in \figurename~\ref{fig:3-spinner gadgets}.

\section{Billiard Balls}
\label{sec:billiard}

Our final application is the billiard ball model, which was introduced by Fredkin and Toffoli \cite{fredkin1982conservative} and is one of the best known reversible models of computation. In the billiard ball model, there are circular \emph{balls} colliding elastically with each other and with fixed \emph{mirrors}. For simplicity, all balls have the same size and mass, and will only move at a single nonzero speed. This model is based on classical physics, and in fact exactly matches the classical kinetic theory of perfect gasses.

The decision problem we consider is whether a ball ever reaches a particular position, given a configuration of mirrors and initial positions and velocities of balls. Fredkin and Toffoli \cite{fredkin1982conservative} proved that this model can perform arbitrary computation by showing how to build and string together Fredkin gates; it follows that the decision problem is PSPACE-complete.

We present a new proof of PSPACE-hardness using our framework. The primary advantage this proof has over Fredkin and Toffoli's is that only a constant number---in particular, two---of balls will be moving at any time, and the two moving balls will always be in close proximity. This means there are fewer details to work out relating to issues like timing; Fredkin and Toffoli had to ensure that signals from disparate parts of the construction arrive at a logic gate simultaneously.

\begin{figure}
  \centering
  \includegraphics[scale=.5]{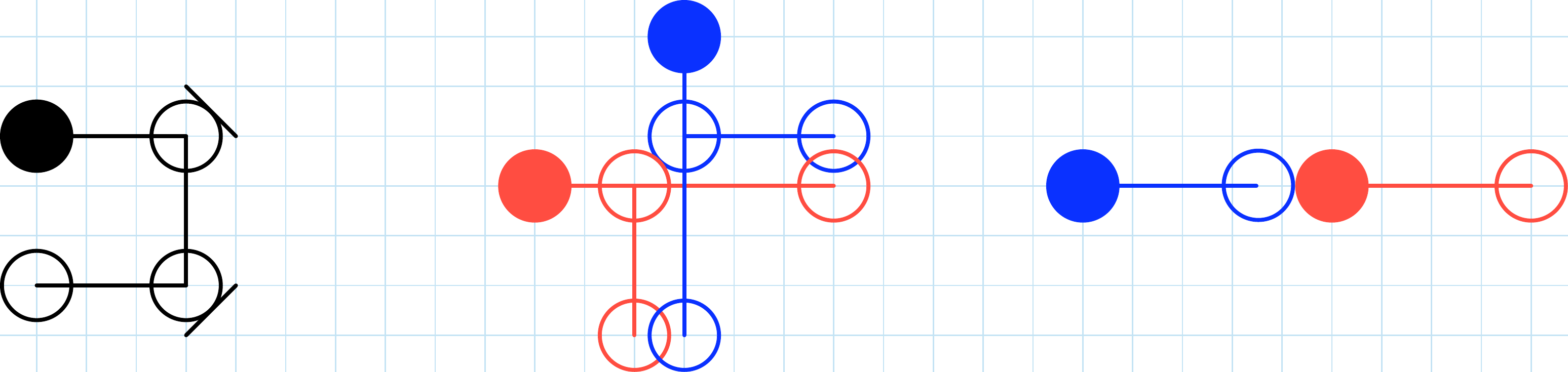}
  \caption{The billiard ball model. Filled circles depict initial positions of balls, and empty circles depict intermediate or final positions. Diagonal lines are mirrors, and horizontal or vertical lines are paths taken by balls. Left: a ball bounces off of mirrors. Middle: two moving balls collide. If only one ball arrives, it goes straight through, but if both balls arrive simultaneously, they bounce off each other. Right: A moving blue ball collides with a stationary red ball, transferring its momentum and leaving the blue ball not grid-aligned.}
  \label{fig:billiard model}
\end{figure}

The balls in our construction all have a radius of $\frac1{\sqrt2}$, and will move only horizontally or vertically. The types of collisions that will occur are shown in \figurename~\ref{fig:billiard model}. One can think of a head-on collision with a stationary ball as moving the stationary ball backwards by the ball diameter, and teleporting the moving ball forwards by the same amount.

\begin{figure}
  \centering
  \includegraphics[scale=.5]{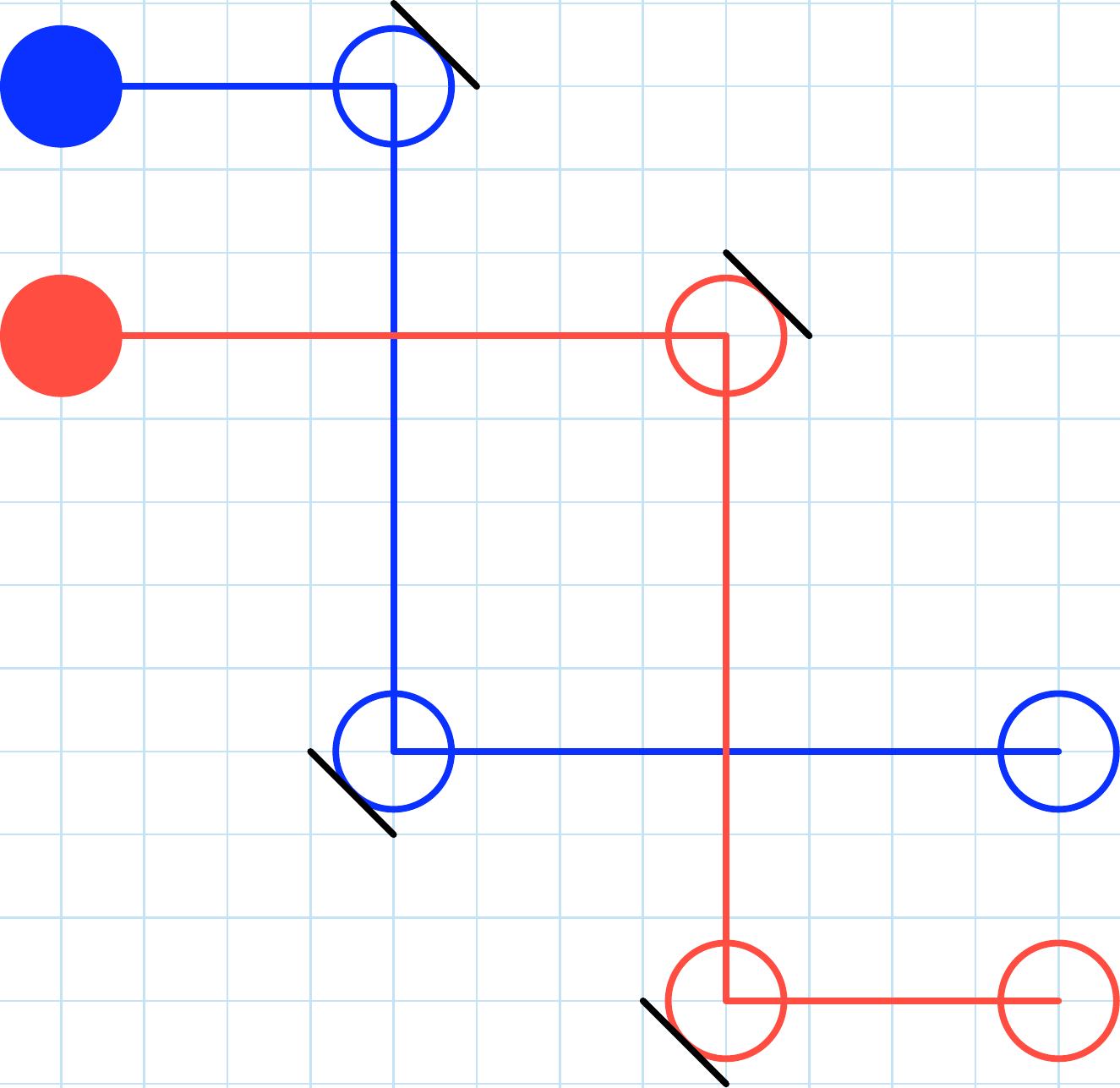}
  \caption{A signal consisting of two billiard balls is sent from the top left to the bottom right. The paths of the two balls have the same length.}
  \label{fig:billiards signal}
\end{figure}

The signal will be represented by two balls moving along parallel paths $2\sqrt2$ (i.e. twice the diameter) apart. This signal is easy to route, as demonstrated by \figurename~\ref{fig:billiards signal}. We will always have the two balls aligned with each other when the signal enters a gadget. Full crossovers, and in particular A/BA crossovers, are trivial: simply have two paths the signal might take cross each other. For simplicity, our diagrams show the paths separated by $3$ units, rather than the actual distance $2\sqrt2\approx2.8$.

\begin{figure}
  \centering
  \includegraphics[scale=.39]{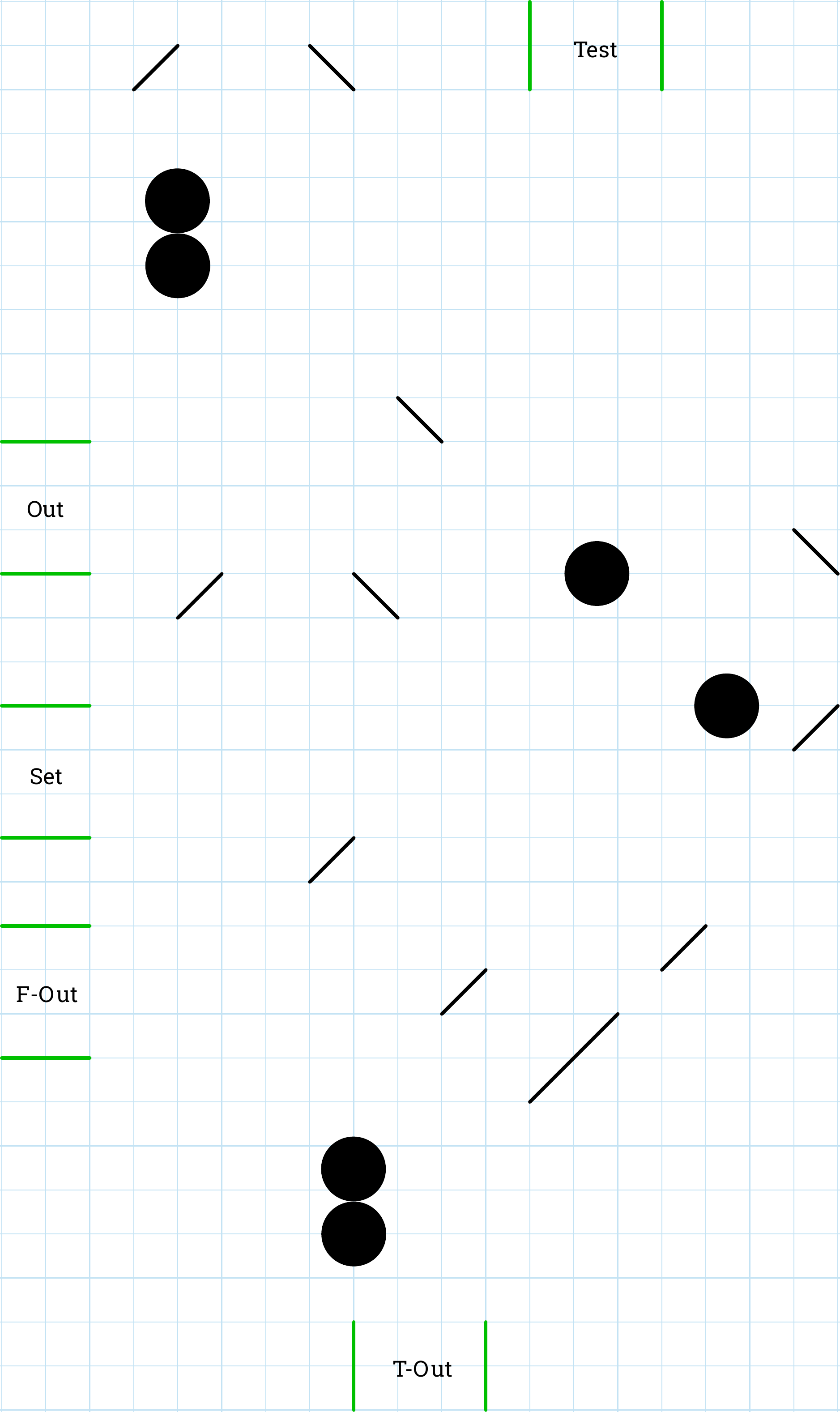}
  \caption{The Switch for the billiard ball model. Each port is marked with a pair of green lines, along which the two balls of the signal may enter or exit.}
  \label{fig:billiards switch empty}
\end{figure}

All that remains is constructing the Switch and Reversible Fan-In. Our Switch is shown in its initial state \figurename~\ref{fig:billiards switch empty}.
The key idea is that stationary balls inside the gadget might (depending on the state) be in the way of one of the balls in the signal entering at Test, effectively making that ball arrive slightly earlier. This change in timing affects whether that ball collides with the other ball in the signal, resulting in two possible places for the signal to end up. 

\begin{figure}
  \centering
  \begin{minipage}[b]{.32\textwidth}
    \includegraphics[scale=.25]{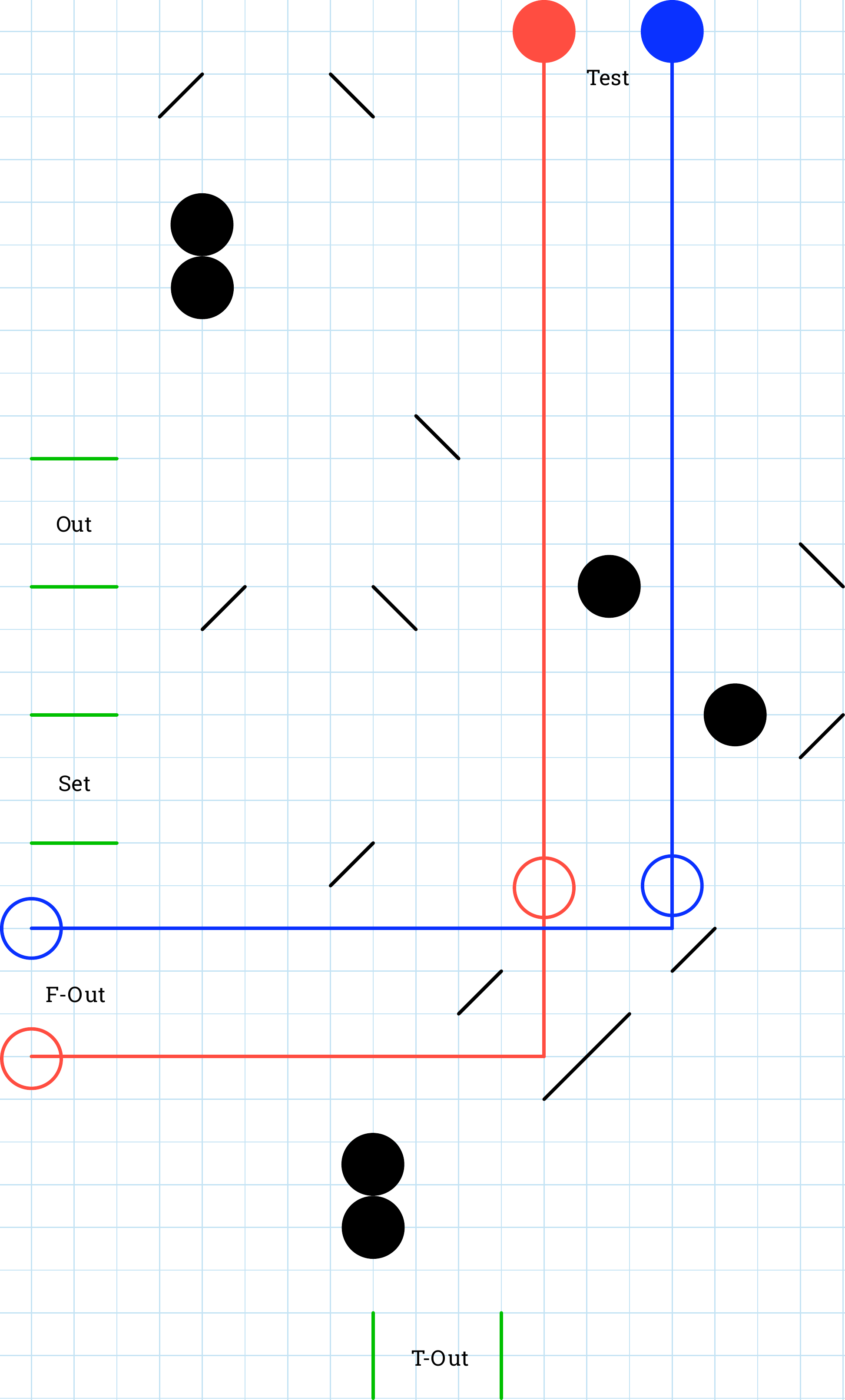}
  \end{minipage}
  \hfill\hfill
  \begin{minipage}[b]{.32\textwidth}
    \includegraphics[scale=.25]{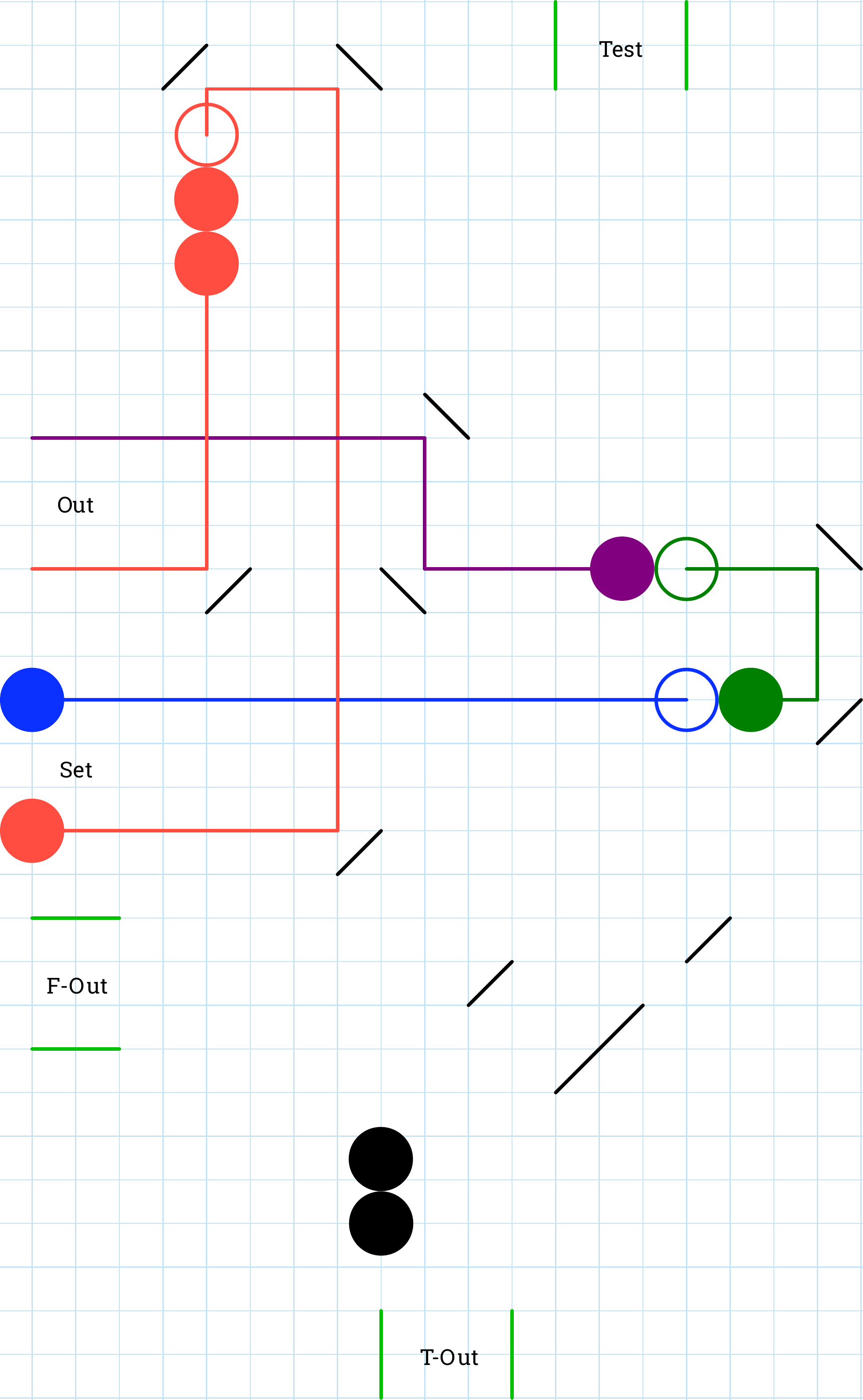}
  \end{minipage}
  \hfill\hfill
  \begin{minipage}[b]{.32\textwidth}
    \includegraphics[scale=.25]{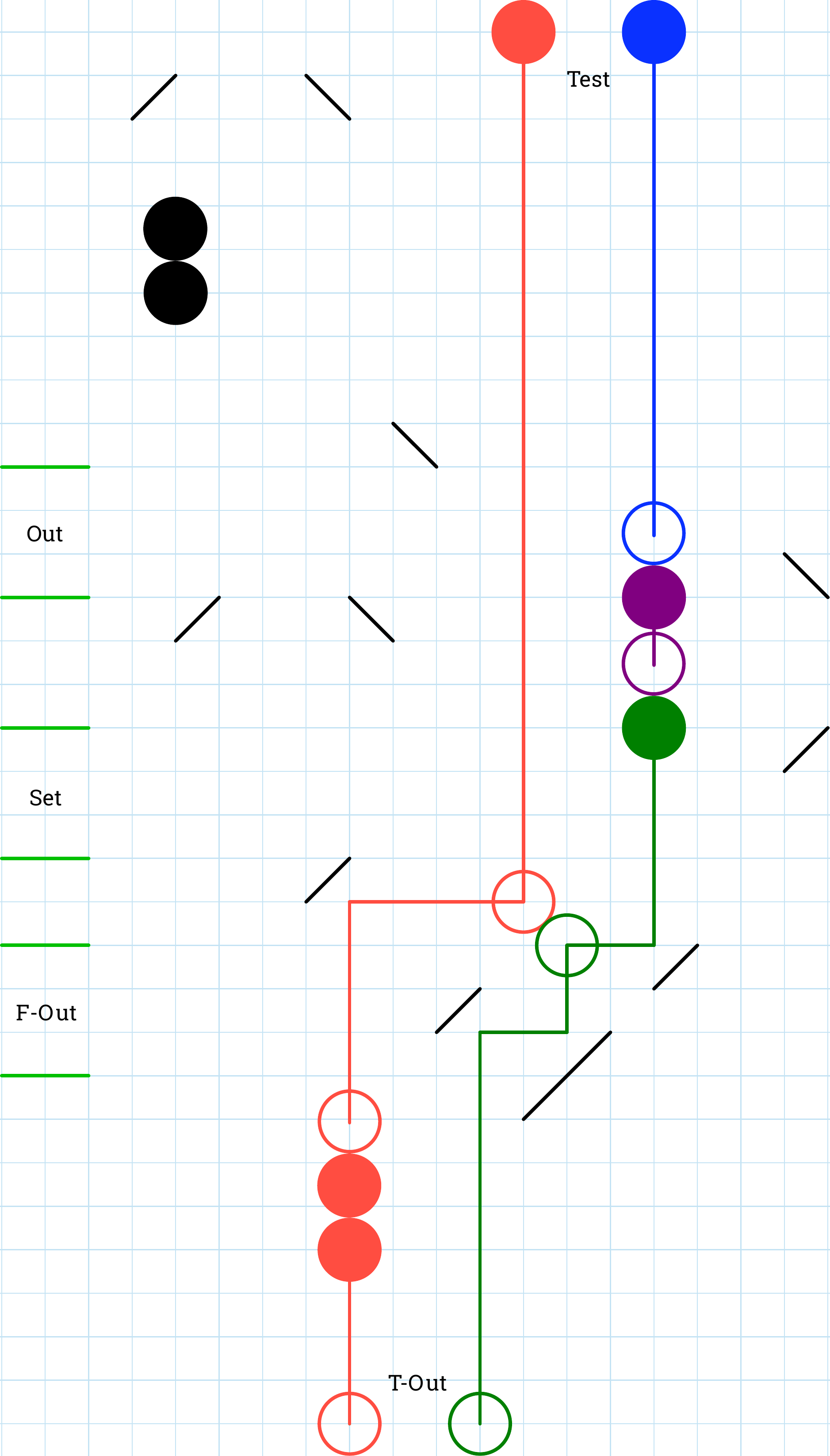}
  \end{minipage}
  \caption{The ways a signal moves through the switch. Left: in the initial state, the signal bounces straight from Test to F-Out. The two balls don't collide where there paths cross. Middle: the blue ball hits the green, which hits the purple, leaving two balls in the path of the Test port. The red ball's path is extended north so that two balls exit at Out simultaneously; the two red balls in its path save the same amount of time as the two balls in the blue ball's path. Right: with the purple and green balls in the way of the signal entering Test, the green ball arrives soon enough to collide with the red ball, resulting in the signal exiting at F-Out. The two additional red balls are again to help synchronize the exit signal.}
  \label{fig:billiards switch}
\end{figure}

The three relevant traversals are shown in \figurename~\ref{fig:billiards switch}.
Since the model has time-reversal symmetry, any gadget built in it also has time-reversal symmetry, so we only need to check that the sequences listed in Table~\ref{fig:gadgets summary} are implemented correctly.

\begin{figure}
  \centering
  \begin{minipage}[b]{.32\textwidth}
    \includegraphics[scale=.25]{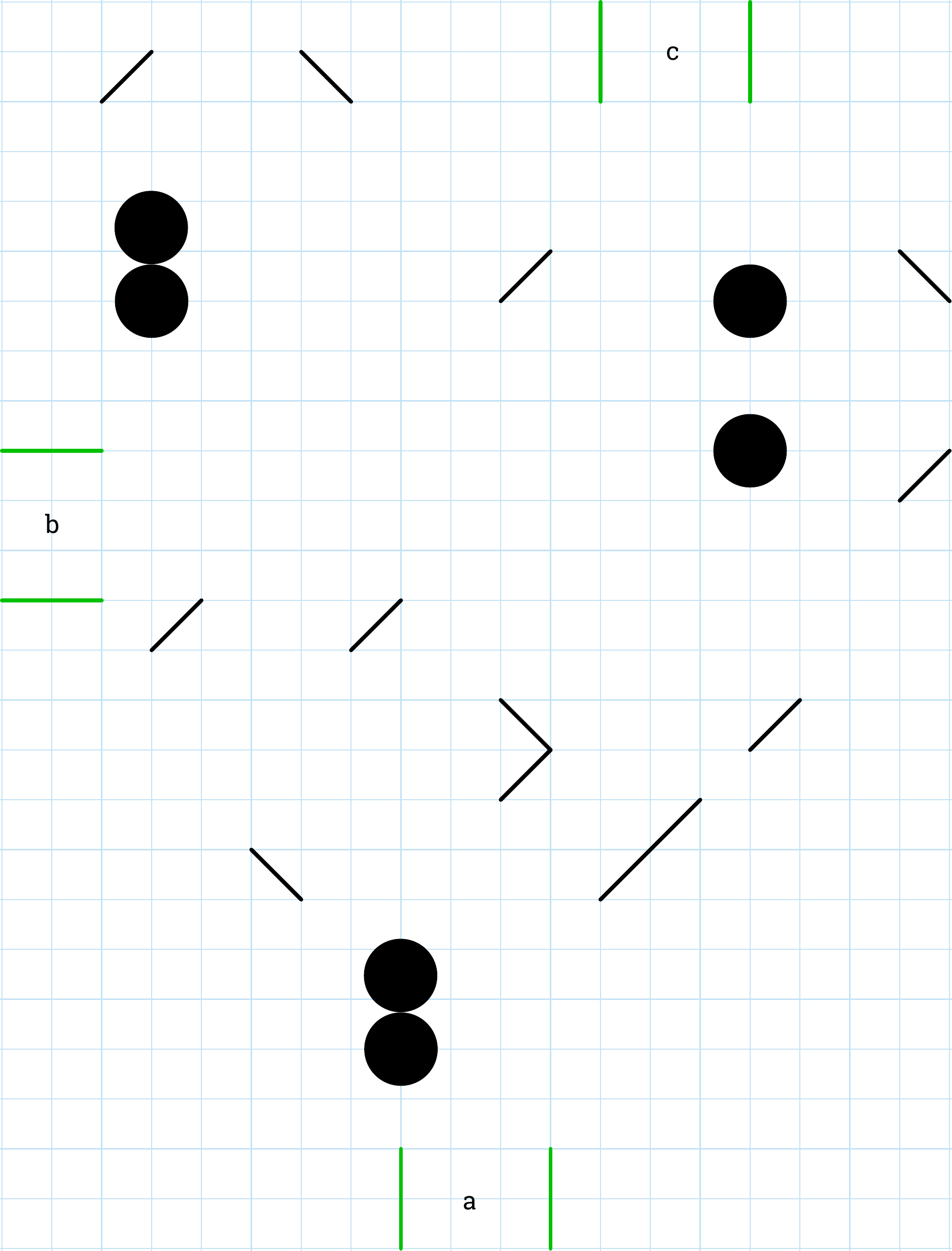}
  \end{minipage}
  \hfill\hfill
  \begin{minipage}[b]{.32\textwidth}
    \includegraphics[scale=.25]{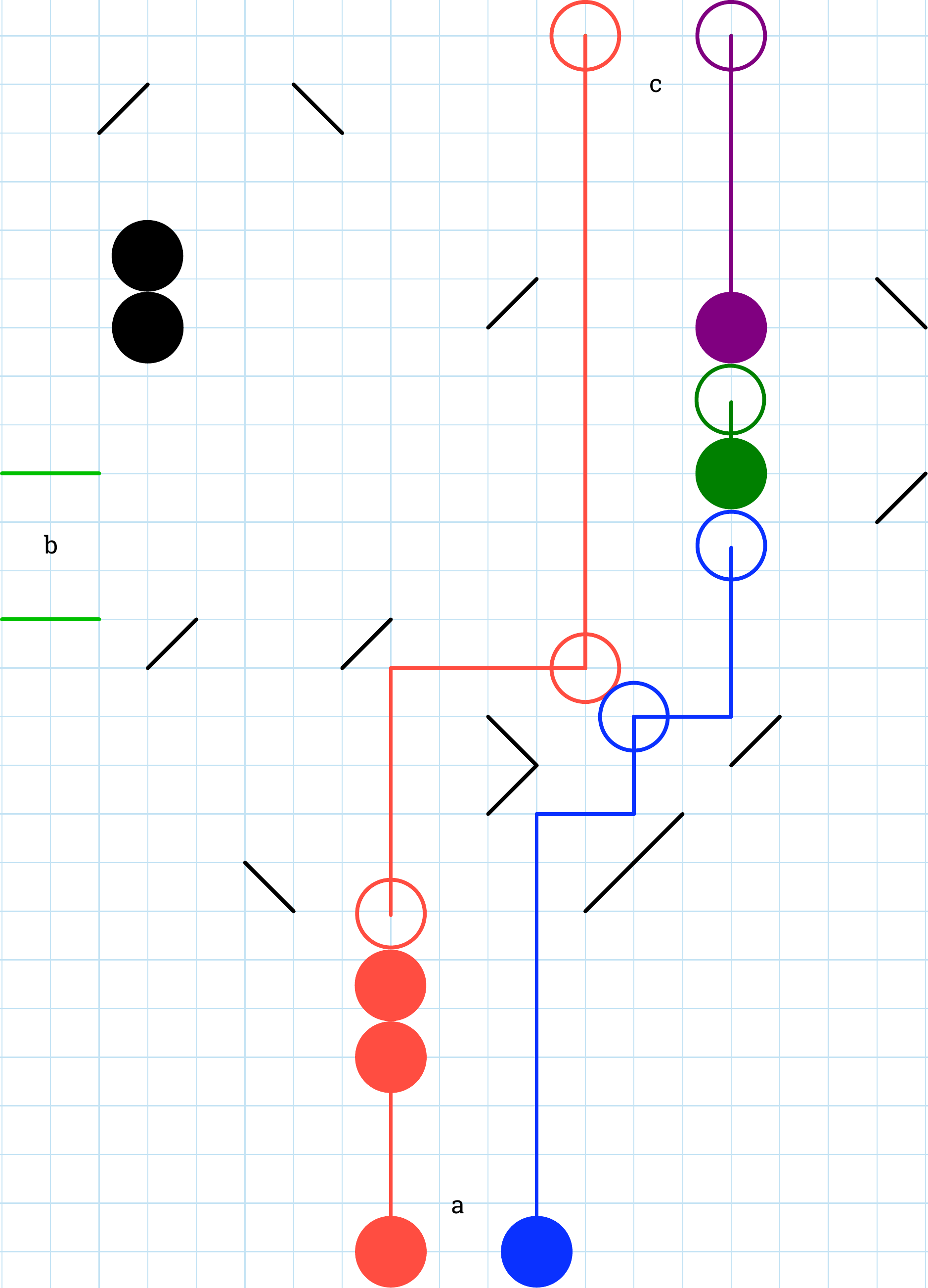}
  \end{minipage}
  \hfill\hfill
  \begin{minipage}[b]{.32\textwidth}
    \includegraphics[scale=.25]{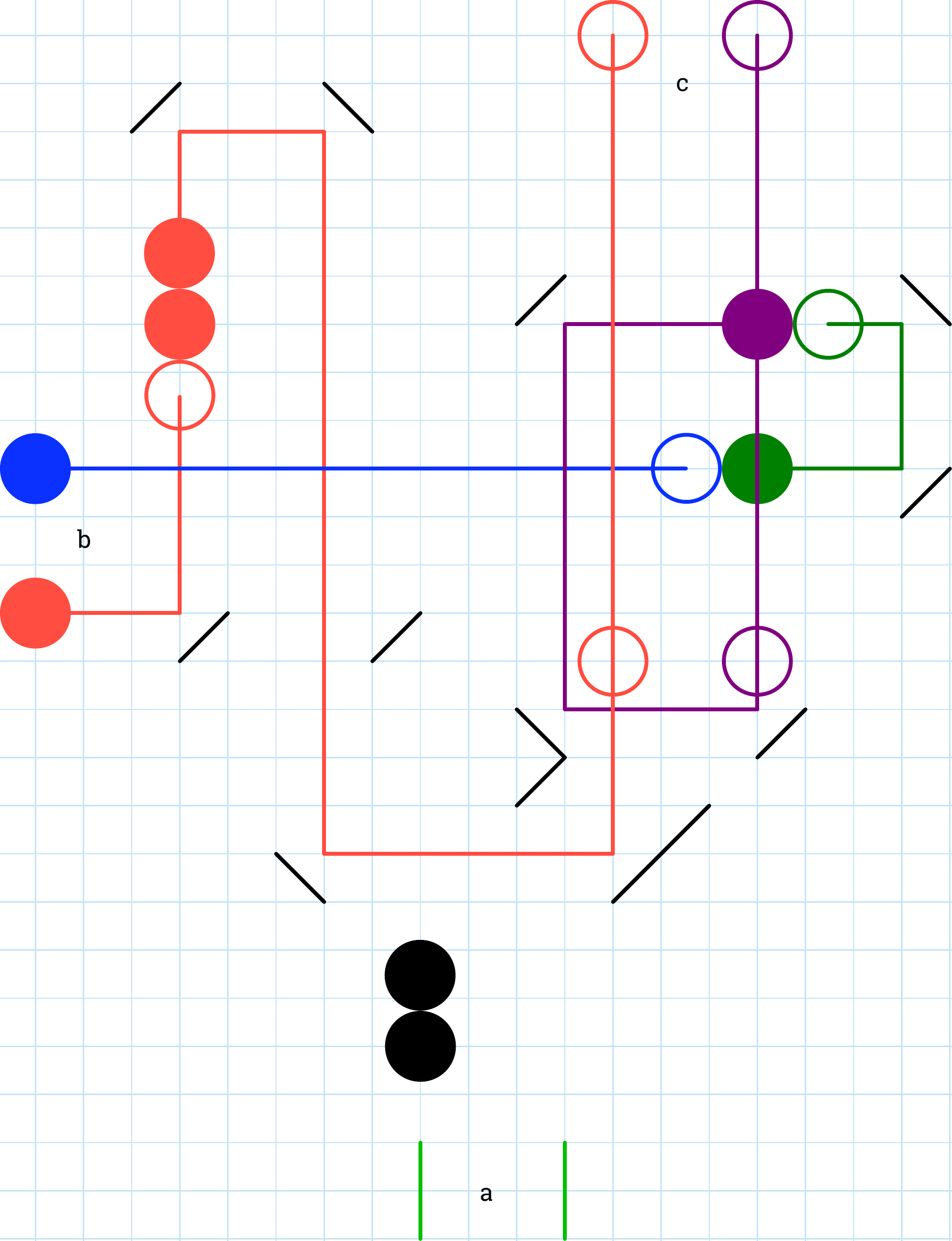}
  \end{minipage}
  \caption{The Reversible Fan-in for the billiard ball model. Left: the gadget in its initial state. Middle: the signal enters at $a$. The signal balls ricochet off each other, and then exit at $c$. They each collide with two stationary balls, so the balls exiting $c$ get there at the same time. Right: The signal enters at $b$. The blue ball knocks the green ball, which knocks the purple ball, clearing the vertical path to $c$. The red ball and the purple ball then exit at $c$ without colliding. The red zigzag to the north and two additional red balls are to make the timing correct.}
  \label{fig:billiards fanin}
\end{figure}

Finally, our Reversible Fan-in is shown in \figurename~\ref{fig:billiards fanin}. It works in a very similar way to Switch, but in reverse, and essentially combining the Set traversal with one of the Test traversals. If the signal enters at $a$, the balls collide and arrive at $c$. If the signal enters at $b$, the signal balls do not collide, and arrive at $c$ with a slightly different timing. To correct the timing, we have the signal entering at $b$ first remove two balls from the path near $c$.

\section{Conclusion}

This paper has analyzed the complexity of three simple models of
reversible deterministic systems: the framework itself,
Deterministic Constraint Logic, and 0-player motion-planning gadgets.
As mentioned in the Introduction, our new results in the last model
can be thought of as extending Table~1 of
\cite{demaine2020toward} to add a `zero-player' column in the unbounded row
(similar to DCL's role in Constraint Logic \cite{hearn2009games}).
These results raise three natural open questions:
\begin{enumerate}
\item Can we fully characterize the complexity of zero-player motion planning
  with a reversible deterministic $k$-tunnel gadget and rotate clockwise, e.g.,
  by showing that the problem is in P --- or even L --- if the gadget does not
  have interacting tunnels? This would nicely complement the analogous
  characterization for one-player motion planning \cite{demaine2020toward}.
\item What about the bounded case: can we characterize the complexity of
  zero-player motion planning with a $k$-tunnel DAG gadget and rotate clockwise?
\item What is the complexity of zero-player motion planning with
  other $2$-state $3$-port reversible deterministic gadgets?
\end{enumerate}

Our proof of PSPACE-hardness of the billiard ball model,
for deciding whether a ball will ever reach a target location,
used at most two balls moving at any time.
Does the problem remain PSPACE-hard for systems
in which only a single ball is moving at any time?

Another direction is analyzing the complexity of other
reversible deterministic systems in computer science,
ideally using the framework or applications of this paper.
One such system to study is
Asynchronous Ballistic Reversible Logic \cite{frank2017asynchronous},
which is asynchronous by removing the ordering assumptions from the model.
Can the ideas from this paper be used to design universal circuits
in this model?
We conjecture that asynchronicity is easy to deal with in our approach
which has only a single signal traveling through the system.

\bibliographystyle{alpha}
\bibliography{main}

\end{document}